\documentclass[12pt]{article}
\usepackage{graphicx}
\usepackage{color}
\usepackage{amssymb}
\usepackage{amsmath}
\def\eq#1{(\ref{#1})}
\def\Eq#1{Eq.~(\ref{#1})}
\newcommand{\secn}[1]{Section~\ref{#1}}

\newcommand{\ket}[1]{|{#1}\rangle}

\def\beq{\begin{equation}}
\def\eeq{\end{equation}}
\def\be{\begin{equation}}
\def\ee{\end{equation}}
\def\beqa{\begin{eqnarray}}
\def\eeqa{\end{eqnarray}}
\def\bea{\begin{eqnarray}}
\def\eea{\end{eqnarray}}

\newcommand{\diag}{\mathrm{diag}\,}
\newcommand{\sect}[1]{\setcounter{equation}{0}\section{#1}}

\renewcommand{\thefootnote}{\fnsymbol{footnote}}

\newcommand{\comm}[2]{\left[#1,#2\right]}
\def\one{{\rm 1\kern -.9mm l}}
\renewcommand{\a}{\alpha}

\newcommand{\ex}[1]{{\rm e}^{#1}}
\def\ii{{\rm i}}
\newcommand{\AB}{A}
\newcommand{\BB}{B}

\newcommand{\GB}{G}
\newcommand{\GCB}{\mathcal{G}}
\newcommand{\EB}{E}
\newcommand{\ECB}{\mathcal{E}}

\newcommand{\RB}{R}

\newcommand{\RCB}{\mathcal{R}}
\newcommand{\SB}{S}
\newcommand{\XB}{X}
\newcommand{\ZB}{Z}
\newcommand{\ZC}{\mathcal{Z}}
\newcommand{\ZCB}{\mathcal{Z}}
\newcommand{\FB}{F}

\newcommand{\FCB}{\mathcal{F}}
\newcommand{\psiB}{\Psi}
\newcommand{\chiB}{\chi}
\newcommand{\ft}[2]{{\textstyle\frac{#1}{#2}}}
\newcommand{\nn}{\nonumber}

\begin{document}
\begin{titlepage}
\hfill SISSA-71/2005/EP
\par\hfill CERN-PH-TH/2005-224
\par\hfill QMUL-PH-05-10
\par\hfill DFTT-05-34

\vspace{15pt}
\begin{center}
{\LARGE \textbf{Brane world effective actions }}

\vspace{20pt}
{\LARGE \textbf{for D-branes with fluxes}}
\end{center}

\vspace{5pt}
\begin{center}
{\bf M. Bertolini}\\
{\sl SISSA/ISAS and INFN - Sezione di Trieste\\
Via Beirut 2; I-34014 Trieste, Italy}

\vspace{5pt}
{\bf M. Bill\`o} \\
{\sl Dipartimento di Fisica Teorica, Universit\`a di Torino \\
and INFN - Sezione di Torino; Via P. Giuria 1; I-10125 Torino, Italy}

\vspace{5pt}
{\bf A. Lerda} \\
{\sl Dipartimento di Scienze e Tecnologie Avanzate,
Universit\`a del Piemonte Orientale \\ Via V. Bellini 25/G; I-15100 Alessandria, Italy\\
and INFN - Sezione di Torino}

\vspace{5pt}
{\bf J.F. Morales and R. Russo\footnote{On leave of absence from {\it Queen Mary,
University of London}, E1 4NS London, UK.}} \\
{\sl CERN, Physics Department\\
CH-1211 Geneva 23, Switzerland}
\end{center}

\begin{center}
\textbf{Abstract}
\end{center}
{\small \noindent We develop systematic string techniques to study
brane world effective actions for models with magnetized (or
equivalently intersecting) D-branes. In particular, we derive the
dependence on all NS-NS moduli of the kinetic terms of the chiral
matter in a generic non-supersymmetric brane configurations with
non-commuting open string fluxes. Near a ${\cal N}=1$ supersymmetric
point the effective action is consistent with a Fayet-Iliopoulos
supersymmetry breaking and the normalization of the scalar kinetic
terms is nothing else than the K\"ahler metric. We also discuss, from
a stringy perspective, $D$ and $F$ term breaking mechanisms, and how,
in this generic set up, the K\"ahler metric enters in the physical
Yukawa couplings.}

\end{titlepage}

\renewcommand{\thefootnote}{\arabic{footnote}}
\setcounter{footnote}{0} \setcounter{page}{1}
\tableofcontents
\vskip 0.8cm
\sect{Introduction and Summary}
\label{intro}

In Type II and Type I string theories, D-branes are the objects
providing, in a simple and natural way, two important features of
our world: the presence of non-abelian gauge groups and that of
four dimensional chiral fermions. In particular, when the ten
dimensional space-time is simply taken to be the direct product of
a six dimensional compact manifold and of a four dimensional
Minkowski part, chiral fermions arise when the D-branes have some
non-trivial properties in the compact space. This can happen when
constant magnetic fields are switched on along the D-brane
world-volume~\cite{Bachas:1995ik}, or when the D-branes intersect
with some non-trivial angles~\cite{Berkooz:1996km} (actually,
these two situations can be usually connected by means of some
T-dualities, see for instance~\cite{Rabadan:2001mt}). By
exploiting these basic features, a new class of string models has
been studied in these last years,
starting from \cite{ Blumenhagen:1999ev,Blumenhagen:2000wh,Angelantonj:2000hi},
providing various interesting phenomenological applications. Recent reviews on this subject,
often named ``Intersecting Brane Worlds''
(IBW)~\cite{Aldazabal:2000cn}, are
Refs.~\cite{Uranga:2003pz,Kiritsis:2003mc,Lust:2004ks,Kokorelis:2004tb,Blumenhagen:2005mu}
and also the detailed derivation of some results can be found in
the
PhD-theses~\cite{MarchesanoBuznego:2003hp,Ott:2003yv,Gorlich:2004zs,Anastasopoulos:2005ba}.
One of the nice features of this class of string models is that
they are ``calculable''. This means that, by using known string
techniques, it is possible to compute explicitly the
Standard-Model-like effective action. Moreover, all the parameters
appearing in such a low-energy action are functions of the
microscopic data specifying the D-brane configuration and the
geometry of the compact space. The explicit derivation of the
effective action is certainly possible whenever the string vacuum
under consideration is described, from the world-sheet point of
view, by some tractable Conformal Field Theory (CFT).  Even if
this is a rather particular set of points in the whole moduli
space of the D-brane/string compactifications, it contains already
some very interesting situations, like those involving orbifolds
or orientifolds and, as we already said, also the case of D-branes
with constant magnetic fields. Thanks to the simplicity of the
underlying string theory, it has been possible to study various
features of the IBW models which go beyond the analysis of the
spectrum and of its quantum numbers. For instance, several authors
studied how the Higgs mechanism~\cite{Cremades:2002cs} and the
Yukawa
couplings~\cite{Cremades:2003qj,Cvetic:2003ch,Abel:2003vv,Lust:2004cx}
are realized in intersecting brane models (or in the T-dual case
of magnetized D-branes~\cite{Cremades:2004wa}); some of their
phenomenological implications are discussed
in~\cite{Abel:2003fk,Abel:2003yh}; threshold
corrections~\cite{Lust:2003ky,Bianchi:2005sa} have been computed;
proton decay can be studied quantitatively~\cite{Klebanov:2003my,Axenides:2003hs};
it has also been shown that the problem of moduli stabilization
can be partly addressed in the framework of solvable string
models, by using D-brane world-volume
fluxes~\cite{Antoniadis:2004pp}. The issue of complete moduli
stabilization has been thoroughly studied, see for instance
Refs.~\cite{Curio:2001qi,Acharya:2002kv,Blumenhagen:2005tn,Bianchi:2005yz,Curio:2005ew,
Villadoro:2005cu,DeWolfe:2005uu,Antoniadis:2005nu,Camara:2005dc,
Lust:2005dy,GarciadelMoral:2005js,Kumar:2005hf,Bianchi:2005sa}; however
generically these constructions go beyond the class of
``solvable'' models we consider in this paper. Various recent
papers discuss phenomenological features of open string models,
where the techniques analyzed in this paper might be useful, see
for instance
Refs.~\cite{Marchesano:2004yq,Ibanez:2004iv,
Marchesano:2004xz,Cvetic:2005bn,Dudas:2005jx,Chen:2005mj,Coriano':2005js}.

In this paper we describe in some generality the string theory
techniques necessary to compute the effective actions for this
class of string models, where the Standard Model fields live on
intersecting or magnetized D-branes. The technique we use is
conceptually simple and well-known: one can reconstruct the
effective action by requiring that it reproduces the low energy
limit of the string amplitudes. Thus this is a two steps
procedure: first it is necessary to compute a string amplitude
contributing to a particular term of the effective action one is
interested in; then one can extract the low-energy amplitude by
sending the string length $\sqrt{\alpha'}$ to zero with all four
dimensional momenta and masses kept fixed. We focus on the dynamics
of the fields coming from the open strings and try to determine the
dependence of the relevant pieces of the four dimensional effective
action on the closed string moduli, whose dynamics is kept frozen ({\i.e.}
we work in a limit where gravity is non dynamical on the brane).
This technique has been explicitly applied in Heterotic string theory by
Dixon, Louis and Kaplunovsky~\cite{Dixon:1989fj} and more recently in the
context of IBW in Ref.~\cite{Lust:2004cx}. Here we follow the same approach,
with the goal to generalize it in various directions. First we show that
this technique is not limited to supersymmetric models.  On the
contrary, it is most effective in situations where supersymmetry
is spontaneously broken, because in this cases we can use the
presence of mass terms to fix unambiguously the overall
normalization of the string amplitudes, which actually plays an
important r\^ole in the form of the resulting effective action.
Then we show that, by using the language of magnetized D-branes,
it is possible to treat in a simple fashion the case of six
dimensional compactifications that are not factorized in products
of two dimensional torii ${\cal T}^2$ (the model discussed in
Ref.~\cite{Antoniadis:2004pp} is in fact already of this type even
if the compact space is ${\cal T}^2\times {\cal T}^2\times {\cal
T}^2$, since the magnetic fields on the brane world-volume do not
respect the factorization of the geometry). In this more generic
situation, contrary to the completely factorized case, the
magnetic fields living on different D-branes do not need to
commute. The presence of non-commuting or oblique fluxes is an
important feature in order to achieve the stabilization of the
off-diagonal moduli in ${\cal T}^6$ (see
Refs.~\cite{Bianchi:2005yz,Bianchi:2005sa} for recent developments
in this direction).  Here we will take also a non-trivial metric
and $B$ field, and show that computations remain manageable even
if the compact space does not have a factorized structure at all.
In a full-fledged model some of the NS-NS moduli are absent due
the presence of orientifolds. However, it is known that these
moduli need not to be trivial, but can be frozen to some non-zero
(discrete) values~\cite{Bianchi:1991eu}. So they will affect the
form of the effective action, and need to be taken into account in
our computations.

As an explicit example of this approach to the derivation of the
effective action, we focus here on the kinetic term for the scalar
fields living at the D-brane intersections. This
term is particularly interesting for two
reasons. In models where we have ${\cal N}=1$ supersymmetry
(possibly spontaneously broken), this term contains the K\"ahler
metric. This function, together with the superpotential and the
normalization of the kinetic terms for the gauge fields, specifies
completely any ${\cal N}=1$ gauge theory
action~\cite{Cremmer:1982en}.  However, in contrast to the other
two building blocks, the K\"ahler metric enters in a
non-holomorphic piece of the action and so has no protection
against string (or quantum) corrections. There is also a stringy
reason that makes the K\"ahler metric interesting. The open
strings stretched between two different D-branes, like those
living at the D-brane intersections, behave like the twisted
sectors of the (Heterotic) orbifold models. This means that the
terms of the effective actions involving this kind of fields
cannot be derived by simple dimensional compactification from the
flat ten dimensional string theory or from Born-Infeld action.
Therefore, the computation of scattering amplitudes represents
basically the only possible way to reconstruct these terms of the
effective action. Our analysis shows that the full (NS-NS) moduli dependence
of the K\"ahler metric is encoded in a disk amplitude with two open
strings and one closed string inserted.

We also consider the scalar fields associated to open strings that
start and end on the same D-brane (corresponding to the string untwisted sector) and compute
their metric. In this case the low-energy dynamics can be readily derived also
from the Born-Infeld action, upon compactification. Then we can check that
the full moduli dependence of the metric for the untwisted fields
is correctly extracted from a three point function
involving two scalars and a generic closed string modulus, showing
the validity of this diagrammatic approach.

\vskip 0.5cm
\subsection{Organization of the paper}

In Section~\ref{secn:modeexpansions} we review the basics of the
open string quantization and this will serve also to set up our
notations. As in the usual case, the open strings stretched
between magnetized or tilted D-branes are more easily analyzed by
using the doubling trick, that is by rewriting the bosonic and
fermionic open string coordinates $x^\mu(\sigma,\tau)$ and
$\chi^\mu(\sigma,\tau)$ in terms of holomorphic CFT's. The
properties of this holomorphic fields depend on the angles or
magnetic fluxes of the D-branes and on the moduli of the compact
space. In particular, in Section~\ref{secn:shifts}, we derive the
relation between the twists $\theta_i$ of the holomorphic fields
and the closed string moduli of the NS-NS sector. We also write
the vertex operators related to these moduli and, in doing so, we
clarify some details about the off-shell continuation of string
amplitudes. In fact this off-shell continuation is necessary, if
one wants to derive the full effective action and not just the
S-matrix elements. In Section~\ref{secn:spectrum} we briefly
review how to derive the open string spectrum for the IBW models
and how to write the vertex operators for open string states. We
also provide a careful analysis of the field theory limit in the
non-supersymmetric case and give the relation between the string
twist parameters $\theta_i$ and the surviving field theory mass
terms. Then, in Section~\ref{secn:kahler}, which contains the main
results of this paper, we compute the dependence of the K\"ahler
metric on the NS-NS moduli. We follow the procedure used in
Ref.~\cite{Lust:2004cx}: we compute a disk amplitude with two open
strings, representing the fields present in the kinetic terms we
are interested in, and a closed string related to a NS-NS modulus.
Clearly this amplitude is related to the {\em variation} of the
quadratic part of the effective action when one of the closed
string moduli is modified and the others are kept fixed. Since the
string computation is exact in all NS-NS parameters, the above
result translates into a differential equation for the K\"ahler
metric.  So we can fix its dependence on the NS-NS v.e.v.'s
exactly to all orders in $\a'$. As anticipated, we consider a
compactification on a generic non-factorized six dimensional
torus, which is equivalent to resum all possible insertions of
soft gravitons in the compact space. Thus our result truly
depends, through the $\theta_i$'s, on all NS-NS moduli, without
any constraint coming from particular hypothesis that are usually
pre-assumed, like the requirement of switching off the moduli
breaking the ${\cal T}^2\times {\cal T}^2\times {\cal T}^2$
factorized structure of the compact space or the supersymmetric
constraints on the $\theta_i$'s~\cite{Lust:2004cx}. Our results
hence generalize (and partially correct, as we shall show)
previous results in the literature. We also discuss, from a string
theory perspective, how supersymmetry breaking is implemented in
these models. We show via a string computation that the masses the
twisted scalars have for generic $\theta_i$ originate from a
Fayet-Iliopoulos term (a D-term supersymmetry breaking). On the
contrary, F-term breaking, which might be present for a generic
choice of the open string fluxes, does not affect the value of
these tree-level masses. Clearly both mechanisms break
supersymmetry in the bulk, too. Previous works on the issue of D
and F term breaking in IBW are
Refs.~\cite{Lust:2004fi,Lust:2004dn,Font:2004cx,Jockers:2005zy}.
Finally, we show that, in the case of factorized fluxes, our
results can be easily translated with three T-dualities in the
Type IIA configuration where the magnetized D9-branes are
described as intersecting D6-branes with generic angles. In
Section~\ref{sec:yukawa}, we discuss Yukawa couplings, focusing on
the quantum (world-sheet) contribution. This part can be
perturbatively expanded in $\alpha'$ and usually does not enter in
the superpotential, which receives only non-perturbative
contributions via world-sheet instantons. This non-renormalization
property~\cite{Witten:1985bz,Dine:1986zy,Dine:1987bq} was proven
in the context of Heterotic models and is certainly interesting to
check whether it holds also in the IBW models. In the factorized
case this non-renormalization property has been checked
in~\cite{Lust:2004cx}, where the authors showed that the quantum
part of the Yukawa couplings can be expressed solely in terms of
the K\"ahler metric. Following our approach, we can prove that the
same property holds in a non-factorized case with commuting
fluxes. In a generic case, the explicit check of the
non-renormalization of the superpotential is difficult, since it
requires to compute a correlator among non abelian twists. Of
course, it is possible to reverse the logic and assume that the
non-renormalization theorem is valid in a general setup also in
open string models. In this case our results provide strong
constraints on the form of the three-point correlator for non
abelian twists which must have a surprisingly simple form.
Appendix \ref{app:jac} contains the derivation of the formula
providing, in a generic situation, the dependence of the open
string twists on the closed string moduli. This enters crucially
in getting the results presented in Section~\ref{secn:kahler}. In
Appendix~\ref{app:unt} we apply exactly the same technique
described in Section~\ref{secn:kahler} to the matter fields
arising from the open strings that start and end on the same
D-brane. In this way we are able to derive the full dependence on
the NS-NS moduli of the Born-Infeld action and of the open string
metric from a disk amplitude with two open strings and one closed
string. This represents a nice test of our diagrammatic approach
to the computation of the low-energy action.

\vskip 0.5cm
\subsection{Outlook}

The main motivation for this work is to provide the techniques to
generalize, in the context of brane world models, previous results
in the literature to potentially more realistic models. Once all
the ingredients to compute effective actions for such a generic
situation are available, it becomes possible to address many
questions in phenomenologically interesting models that have been
recently constructed. In this respect, string theory techniques
prove to be an efficient tool to compute low energy effective
actions whenever this cannot be done otherwise. There are,
however, a number of issues we have not addressed in this work and
which we leave to future investigations. The most technically
difficult but interesting thing to do would be to compute directly
the Yukawa three-point function in the generic case, {\it i.e.}
for non-abelian twists. Moreover, in our string computations, we
neglect the contribution coming from world-sheet instantons; it is
of course very important to include them systematically in our
approach. We have not discussed the dependence on the R-R moduli,
but these should be included in a complete low energy effective
description. Similarly, we have not considered open string moduli
({\em i.e.} Wilson lines), whose stabilization, in IBW models, has
not been addressed in much detail, so far. Finally, the K\"ahler
potential is a D term hence is not protected by
non-renormalization theorems and would then be very interesting to
compute higher loop corrections in the string coupling (recent
results in this direction can be found in
Refs.~\cite{Berg:2005ja,Abel:2004ue,Abel:2005qn}).

\vskip 0.8cm
\sect{Open Strings in Closed String Background}
\label{secn:modeexpansions}

In this section we review the quantization of open strings moving in a
$2d$-dimensional Euclidean space with a constant metric $G$ and a
constant NS-NS antisymmetric tensor $B$. This case is relevant in
discussing systems of magnetized D-branes or, after T-dualities,
systems of intersecting D-branes.

\vskip 0.5cm
\subsection{Bosonic sector}
We begin our analysis by considering the bosonic sector described by
the string coordinates $x^M$ ($M=1,...,2d)$ whose action (in a
Euclidean world-sheet) is
\begin{equation}
\label{Sx}
S_{\rm bos}  =
-\frac{1}{4\pi\alpha'}\!\int \!d^2\xi\,\Big[
\partial^\alpha x^M \partial_\alpha x^N G_{MN} +
\ii \epsilon^{\alpha\beta} \partial_\alpha x^M \partial_\beta x^N
B_{MN}\Big]
-\, \ii \sum_{\sigma} q_{\sigma} \!\int_{C_{\sigma}} \!\!\!\!
dx^M A^{\sigma}_M\,,
\end{equation}
where the index $\sigma$ on $C$, $A$ and $q$ takes the values
$\sigma=0$ or $\sigma=\pi$ and labels the string end-points;
$q_{\sigma}$ is the charge with respect to a background gauge field
$\AB^{\sigma}$ along the boundary $C_{\sigma}$. Our conventions are
such that $q_\pi=-q_0=1$ and $\epsilon^{\sigma\tau}=1$; the string
coordinates $x^M$ and the gauge fields $\AB^\sigma$ are dimensionless,
while the background metric $\GB$ and the $\BB$ field have dimensions
of (length)$^2$. In the following we will consider only the case in
which $\GB$ and $\BB$ are constant, and the gauge fields
$\AB_{\sigma}$ are linear with constant field strengths
$\FB_{\sigma}$. Then, it is easy to realize that the field equations
$\partial^\alpha\partial_\alpha x^M=0$ must be supplemented by the
following boundary conditions
\begin{equation}
\Big(G_{MN}\partial_\sigma x^N+{\rm i}({\cal F}_\sigma)_{MN}\partial_\tau
x^N\Big)\Big|_{\sigma=0,\pi}=0~,
\label{bc}
\end{equation}
where
\begin{equation}
\FCB_{\sigma}
=\BB + 2\pi\alpha' \,\FB_{\sigma}~.
\label{calF}
\end{equation}
Introducing the complex variable $z={\rm e}^{\tau+{\rm i}\sigma}$ and the reflection matrices
\begin{equation}
\RB_\sigma=\big(\GB-\FCB_\sigma\big)^{-1}\,\big(\GB+\FCB_\sigma\big)~~,
\label{R}
\end{equation}
the boundary conditions (\ref{bc}) can be rewritten
as
\begin{equation}
\overline\partial x^M\Big|_{\sigma=0,\pi}=(R_\sigma)^M_{~N}\,\partial
x^N\Big|_{\sigma=0,\pi}~~.
\label{bc1}
\end{equation}

A convenient way to solve these equations is to
define, in the complex $z$-plane, multi-valued chiral fields $X^M(z)$ such that
\begin{equation}
X^M({\rm e}^{2\pi{\rm i}}z) =
\big(R_\pi^{-1}\,R_0\big)^{M}_{~N}\,X^N(z) \equiv R^{M}_{~N}\,X^N(z)
~,~~\mbox{where}~~~~\RB \equiv \RB_\pi^{-1} \RB_0~~.
\label{chiraly}
\end{equation}
Then, putting the branch cut in the $z$-plane just below the negative real axis,
a solution to the boundary conditions (\ref{bc1}) is
\begin{equation}
x^M(z,\overline z) = q^M+ \frac{1}{2}\Big[X^M(z) +\big(R_0\big)^{M}_{~N}\,X^N(\overline
z)\Big]~~,
\label{sol}
\end{equation}
where $z$ is restricted to the upper half-complex plane, and $q^M$
are constant zero-modes.

Let us observe that the reflection matrix $R_\sigma$ defined in
(\ref{R}) leaves the metric $G$ invariant:
\begin{equation}
{}^t \RB_\sigma
\,\GB\,\RB_\sigma=\GB ~,
\label{orthogonality}
\end{equation}
and so does, as a consequence, the monodromy matrix $\RB$.  Then,
introducing the vielbein $E_{~M}^A$, such that $G_{MN} =
E_{~M}^A E_{~N}^B \delta_{A B}$, we see that in the new basis $R_{AB}$ is
simply a $\mathrm{SO}(2d)$ matrix, so that it is always possible to
find an orthonormal frame and a unitary transformation to put the
monodromy matrix in a diagonal form, namely
\begin{equation}
\ECB\,\RB\,\ECB^{-1}
\equiv \RCB
={\rm diag}\Big(
{\rm e}^{2{\rm i}\pi\theta_1},\cdots ,{\rm e}^{2{\rm i}\pi\theta_d},
{\rm e}^{-2{\rm i}\pi\theta_1},{\rm e}^{-2{\rm
i}\pi\theta_d}\Big)
\label{diagonalR1}
\end{equation}
for $0\leq\theta_i<1$. Let us point out that in the resulting complex
basis $\ZCB = \big(\ZC^i\,,\,\bar\ZC^i\big)$ given by $\ZCB = \ECB
\XB$, the metric is
\begin{equation}
\label{Gcal}
\GCB = {}^{\rm t}{\ECB}^{-1}\,\GB\,{\ECB}^{-1}
=\left(\begin{matrix}
0 & \one\\
\one & 0
\end{matrix}
\right)
\end{equation}
and the monodromy properties (\ref{chiraly}) become
\begin{equation}
\mathcal{Z}^{\,i}({\rm e}^{2\pi{\rm i}}z) = {\rm e}^{2{\rm
i}\pi\theta_i}\,\mathcal{Z}^{\,i}(z)~~~~{\rm and}~~~~
\overline{\mathcal{Z}}^{\,i}({\rm e}^{2\pi{\rm i}}z) = {\rm e}^{-2{\rm
i}\pi\theta_i}\,\overline{\mathcal{Z}}^{\,i}(z)
\label{calY1}
\end{equation}
for $i=1,...,d$. Upon canonical quantization, we obtain the following mode
expansions\footnote{We have included appropriate prefactors to recover the
standard expansions for $\theta_i=0$ of dimensionful string fields $\mathcal{Z}^i$.
This means that the matrix $\ECB$ of the change of basis
has dimensions of (length).}:
\begin{subequations}
\label{modexpansionsZ}
\begin{align}
&\partial\mathcal{Z}^{\,i}(z) = -{\rm i}\,\sqrt{2\alpha'}\Bigg(
\sum_{n=1}^\infty\overline{a}^i_{n-\theta_i}\,z^{-n+\theta_i-1}
+\sum_{n=0}^\infty{a}^{\dagger\,i}_{n+\theta_i}\,z^{n+\theta_i-1}
\Bigg)~~,
\label{y1}\\
&\partial\overline{\mathcal{Z}}^{\,i}(z) = - {\rm
i}\,\sqrt{2\alpha'}\Bigg(
\sum_{n=0}^\infty {a}^i_{n+\theta_i}\,z^{-n-\theta_i-1}
+\sum_{n=1}^\infty {\overline a}^{\dagger\,i}_{n-\theta_i}
\,z^{n-\theta_i-1} \Bigg)~~.
\label{ybar1}
\end{align}
\end{subequations}
We remark that the shifts $\theta_i$ are related to the eigenvalues of
the monodromy matrix $\RB$, and not to the two individual reflection
matrices $\RB_0$ and $\RB_\pi$, which, in general, do not commute with
each other. If one or more $\theta_i$'s are zero, particular care must
be paid due to the appearance of extra zero-modes in the corresponding
chiral bosons. In the following, however, we will consider the generic
case in which all shifts are non-vanishing. Canonical quantization
implies that also the zero-modes $q^M$ in~\eq{sol} are operators that
do not commute among them~\cite{Abouelsaood:1986gd}. This fixes the
degeneracy of the open string vacuum, but we will not need this
information in what follows.

The modes appearing in (\ref{modexpansionsZ}) obey the following
commutation relations
\begin{subequations}
\begin{align}
&\big[\overline{a}^i_{n-\theta_i}\,,\,\overline{a}^{\dagger\,j}_{m-\theta_j}\big]
=(n-\theta_i)\,\delta^{ij}\,\delta_{n,m}~~~~~~\forall n,m\geq 1~~,
\\
&\big[{a}^i_{n+\theta_i}\,,\,{a}^{\dagger\,j}_{m+\theta_j}\big]
=(n+\theta_i)\,\delta^{ij}\,\delta_{n,m}~~~~~~\forall n,m\geq0~~.
\end{align}
\label{comm}
\end{subequations}
In particular the oscillators
$\overline{a}^i_{n-\theta_i}$ and
${a}^i_{n+\theta_i}$ are annihilation operators, whereas ${\overline a}^{\dagger\,i}_{n-\theta_i}$
and ${a}^{\dagger\,i}_{n+\theta_i}$ are the corresponding creation operators
with respect to the {twisted} vacuum
$|\Theta\rangle\equiv|\{\theta_i\}\rangle$, {\it i.e.} for any $i$
\begin{equation}
\overline{a}^i_{n-\theta_i}|\Theta\rangle=0~~~\forall \,n\geq 1
~~~~{\rm and}~~~~{
a}^{i}_{n+\theta_i}|\Theta\rangle=0~~~\forall \,n\geq
0~~.
\label{annihil}
\end{equation}

The contribution of the bosons ${\cal Z}^i$ to the
Virasoro generators can be easily derived from the action
(\ref{Sx}), and in particular one finds that
\begin{equation}
L^{({\cal Z})}_0 =\sum_{i=1}^d\Bigg[\sum_{n=1}^\infty
\overline{a}^{\dagger\,i}_{n-\theta_i} \overline{a}^i_{n-\theta_i}
+\sum_{n=0}^\infty {a}^{\dagger\,i}_{n+\theta_i}{a}^{i}_{n+\theta_i}
+\frac{1}{2}\theta_i(1-\theta_i)\Bigg]~=~N^{({\cal Z})} + c^{({\cal Z})}~,
\label{L0}
\end{equation}
where in the last step we have distinguished the operator
$N^{({\cal Z})}$ which measures the number of $a$ and $\overline a$
oscillators from the c-number $c^{({\cal Z})}$ due to the normal ordering with respect
to the twisted vacuum introduced above. From this expression, we can see that
$|\Theta\rangle$ is related to the $\mathrm{Sl}(2,\mathbb{R})$ invariant vacuum $|0\rangle$
through the action of $d$ twist fields~\cite{Dixon:1986qv} $\sigma_{\theta_i}(z)$
of conformal dimensions
\begin{equation}
h_{\sigma_{\theta_i}}=\frac{1}{2}\,\theta_i(1-\theta_i)
\label{hsigma}
\end{equation}
as follows
\begin{equation}
|\Theta\rangle = \lim_{z\to 0}\,\,\prod_{i=1}^d
\sigma_{\theta_i}(z)\,|0\rangle~~.
\label{vacuum}
\end{equation}
On the other hand, the conjugate vacuum $\langle -\Theta|$ is
obtained by acting at infinity with the conjugate twist fields
$\sigma_{-\theta_i}(z)$ of conformal dimensions $h_{\sigma_{-\theta_i}}
=h_{\sigma_{\theta_i}}$,
namely
\begin{equation}
\langle -\Theta| = \lim_{z\to\infty}\,\langle 0|\,\prod_{i=1}^d
\Big(\sigma_{-\theta_i}(z)\, z^{2h_{\sigma_{-\theta_i}}}\Big)~~.
\label{bravacuum}
\end{equation}
Normalizing the vacuum states in such a way that $\langle
-\Theta|\Theta\rangle=1$, from (\ref{vacuum}) and
(\ref{bravacuum}) it immediately follows that
\begin{subequations}
\begin{align}
& \sigma_{-\theta_i}(z)\,\sigma_{\theta_j}(w)\,\sim
\,\frac{\delta^{ij}}{(z-w)^{\theta_i(1-\theta_i)}}~~~~,~~
&\partial\mathcal{Z}^{\,i}(z)\,\partial\overline{\mathcal{Z}}^{\,j}(w)
\sim -\,\frac{{2\alpha'\, \delta^{ij}}}{\left(z-w\right)^2}~~,
\label{freeprop}
 \\
& \frac{\partial {\cal Z}^{\,i}(z)}{\sqrt{2\a'}}\,\sigma_{\theta_j}(w)\sim
\frac{\delta^{ij}\;\tau_{\theta_j}(w)}{(z-w)^{1-\theta_j}}~~~~, ~~
&\frac{\partial \overline{\cal Z}^{\,i}(z)}{\sqrt{2\a'}}\,\sigma_{\theta_j}(w)\sim
 \frac{\delta^{ij}\;\overline{\tau}_{\theta_j}(w)}{(z-w)^{\theta_j}}~~.
\label{OPEsigma}
\end{align}
\end{subequations}
Exploiting the mode expansions (\ref{modexpansionsZ}) and the properties of the twisted
vacuum, it is straightforward to show that
\begin{equation}
\langle
-\Theta|\,\partial\mathcal{Z}^{\,i}(z)\,\partial\overline{\mathcal{Z}}^{\,j}(w)
\,|\Theta\rangle
=-\left(\frac{w}{z}\right)^{-\theta_i}
\frac{{2\alpha'}\,\delta^{ij}}{(z-w)^2}\left[1-\theta_i\left(1-\frac{w}{z}\right)\right]~.
\label{2point}
\end{equation}
Then, by performing a projective $\mathrm{Sl}(2,\mathbb{R})$
transformation, we can move the position of the twist fields to
arbitrary positions and obtain
\begin{eqnarray}
A_{\rm bos}^i(z_1,...,z_4|\sigma_{\theta_i}) &\equiv& \frac{\big\langle \sigma_{-\theta_i}(z_1)\,
\partial\mathcal{Z}^{\,i}(z_2)\,\partial\overline{\mathcal{Z}}^{\,i}(z_3)
\,\sigma_{\theta_i}(z_4)\big\rangle}{\big\langle \sigma_{-\theta_i}(z_1)\,
\,\sigma_{\theta_i}(z_4)\big\rangle
\,\big\langle
\partial\mathcal{Z}^{\,i}(z_2)\,\partial\overline{\mathcal{Z}}^{\,i}(z_3)\big\rangle}
\nonumber \\
&=&
{\omega^{-\theta_i}}
\,\Big[1-\theta_i(1-\omega)\Big]~,
\label{boscorr}
\end{eqnarray}
where $\omega$ is the anharmonic ratio
\begin{equation}
\omega = \frac{\left(z_1-z_2\right)\left(z_3-z_4\right)}{\left(z_1-
z_3\right)\left(z_2-z_4\right)}~~.
\label{anharmonic}
\end{equation}
It is interesting to observe that
\begin{equation}
\big\langle \sigma_{-\theta_i}(z_1)\,
\partial\mathcal{Z}^{\,i}(z_2)\,\partial\overline{\mathcal{Z}}^{\,i}(z_3)
\,\sigma_{\theta_i}(z_4)\big\rangle
=\big\langle \sigma_{-\theta_i}(z_1)\,\partial\overline{\mathcal{Z}}^{\,i}(z_3)
\,\partial\mathcal{Z}^{\,i}(z_2)
\,\sigma_{\theta_i}(z_4)\big\rangle
\label{boscorr1}
\end{equation}
which can be proved with an explicit calculation along the same lines
outlined above.

\vskip 0.5cm
\subsection{Fermionic sector}
Let us now turn to the fermionic sector described by world-sheet
spinors $\chi^M$, whose Euclidean world-sheet action is \cite{Haggi-Mani:2000uc}
\begin{equation}
S_{\rm ferm}=-\frac{{\rm i}}{4\pi\alpha'}\!\int\! d^2\xi
~{\overline\chi}^M\rho^\alpha\partial_\alpha\chi^N\big(G_{MN}+B_{MN}\big)
-\frac{\rm i}{2}\,
\sum_{\sigma} q_{\sigma} \!\int_{C_{\sigma}} \!\!\!\!
d\tau\,{\overline\chi}^M\rho^\tau \chi^N F_{MN}^\sigma~,
\label{Sferm}
\end{equation}
where $\rho^\alpha$ are the 2-dimensional Dirac matrices. Denoting
by $\chi_-^M$ and $\chi_+^M$, respectively, the upper and lower components of
$\chi^M$, from the above action one finds that the standard
field equations $\partial_\pm\chi^M_\mp=0$ must be supplemented by
the following boundary conditions
\begin{equation}
\chi_-^M\Big|_{\sigma=0}=(R_0)^M_{~N}\,\chi_+^N\Big|_{\sigma=0}
~~~~{\rm and}~~~~
\chi_-^M\Big|_{\sigma=\pi}=-\eta\,(R_\pi)^M_{~N}\,\chi_+^N\Big|_{\sigma=\pi}~,
\label{bcferm}
\end{equation}
where $\eta=1$ for the NS sector and $\eta=-1$ for the R sector.
The solution to these equations can be conveniently written in
terms of multi-valued chiral fermions $\psi^M(z)$ such that
\begin{equation}
\psi^M({\rm e}^{2\pi{\rm i}}z) =\eta\,
\big(R_\pi^{-1}\,R_0\big)^{M}_{~N}\,\psi^N(z)~.
\label{chiralfer}
\end{equation}
Indeed, remembering that the fermionic fields have conformal dimensions $1/2$, we have
\begin{equation}
\psi_+^M(z)\equiv z^{\,-1/2}\,\chi_+^M(z)=\psi^M(z)~~~~{\rm and}~~~~
{\psi}^M_-(\overline z)\equiv{\overline z}^{\,-1/2}\,\chi_-^M(\overline z)
=\big(R_0\big)^{M}_{~N}\,\psi^N(\overline{z})
\label{psiz}
\end{equation}
for any $z$ with ${\rm Im}\,z\geq 0$. In the complex basis $\psiB = \ECB\,\chiB=
\big(\Psi^{\,i}\,,\,\overline{\Psi}^{\,i}\big)$, where the monodromy
matrix $\RB$ is diagonal, we can rewrite (\ref{chiralfer}) simply as
\begin{equation}
\Psi^{\,i}({\rm e}^{2\pi{\rm i}}z)= \eta\,{\rm e}^{2\pi{\rm
i}\theta}\,\Psi^{\,i}(z)~~~~{\rm and}~~~~
{\overline\Psi}^{\,i}({\rm e}^{2\pi{\rm i}}z)= \eta\,{\rm e}^{-2\pi{\rm
i}\theta}\,{\overline\Psi}^{\,i}(z)
\label{fermionitwist}
\end{equation}
and, after canonical quantization, obtain the following mode
expansions
\begin{subequations}
\label{fermexpansions}
\begin{align}
&\Psi^{\,i}(z) = \sqrt{2\alpha'}\sum_{n=0+\nu}^\infty\Big(
{\overline \Psi}^{\,i}_{n-\theta_i}\,z^{-n+\theta_i-\frac{1}{2}}
+{\Psi}^{\,\dagger \,i}_{n+\theta_i}\,z^{n+\theta_i-\frac{1}{2}}\Big)
\label{expansionNS1}\\
&{\overline \Psi}^{\,i}(z)=\sqrt{2\alpha'}\sum_{n=0+\nu}^\infty\Big(
{\Psi}^{\,i}_{n+\theta_i}\,z^{-n-\theta_i-\frac{1}{2}}
+{\overline \Psi}^{\,\dagger \,i}_{n-\theta_i}\,z^{n-\theta_i-\frac{1}{2}}\Big)~,
\label{expansionNS2}
\end{align}
\end{subequations}
where $\nu=0$ in the R sector, $\nu=1/2$ in NS sector. The modes in
(\ref{fermexpansions}) obey the following anticommutation relations
\begin{equation}
\big\{\Psi^{\,i}_{n+\theta_i}\,,\,\Psi^{\,\dagger \,i}_{m+\theta_i}\big\}
\,=\, \big\{{\overline
\Psi}^{\,i}_{n-\theta_i}\,,\,{\overline\Psi}^{\,\dagger\,i}_{m-\theta_i}\big\}
\,=\,\delta^{ij}\,\delta_{n,m}~~~~~~\forall \,n,m\geq 0+\nu~~.
\label{anticomm}
\end{equation}

Let us concentrate on the NS sector ($\nu=1/2$).  The oscillators
$\Psi^{\,i}_{n+\theta_i}$ and ${\overline\Psi}^{\,i}_{n-\theta_i}$ are
annihilation operators, while $\Psi^{\,\dagger\, i}_{n+\theta_i}$ and
${\overline\Psi}^{\,\dagger\, i}_{n-\theta_i}$ are creation operators
with respect to the fermionic twisted NS vacuum $|\Theta\rangle_{\rm
  NS}$, {\it i.e.}
\begin{equation}
\Psi^{\,i}_{n+\theta_i}|\Theta\rangle_{\rm NS}\,=\,
{\overline \Psi}^{\,i}_{n-\theta_i}|\Theta\rangle_{\rm NS}\,=\,0
~~~~~~\forall \,n\geq \frac{1}{2}~~.
\label{fermvacuum}
\end{equation}
Notice that this definition of creation/destruction operators is
natural only for $0 \leq \theta_i < \frac 12$. In this range, in fact,
the oscillator ${\overline\Psi}_{\frac 12 -\theta_i}^{\,\dagger\, i}$
is a true creation operator since it increases the energy by the
positive amount $\big(\frac 12 - \theta_i\big)$. If, instead, $\frac
12 < \theta_i < 1$, the oscillator ${\overline\Psi}_{\frac
  12-\theta_i}^{\,\dagger \,i}$ {decreases} the energy of the state it
acts on. Thus, in this case the roles of the NS vacuum
$|\Theta\rangle_{\rm NS}$ and of ${\overline\Psi}_{\frac
  12-\theta_i}^{\,\dagger\, i} |\Theta\rangle_{\rm NS}$ are exchanged
and the latter state becomes the true vacuum of the theory, since it
has lower energy. This will be relevant in the discussion of the GSO
projection, see Section~\ref{secn:spectrum}.

{F}rom the action (\ref{Sferm}), one can easily derive the fermionic
contribution to the Virasoro generators, and in particular one finds
\begin{equation}
L_0^{(\Psi)}= \sum_{i=1}^d\! \Big\{\!
\sum_{n=\frac{1}{2}}^\infty\!\Big[(n+\theta_i)
\Psi^{\,\dagger \,i}_{n+\theta_i}\Psi^{\, i}_{n+\theta_i}
\!+(n-\theta_i)
{\overline \Psi}^{\,\dagger \,i}_{n-\theta_i}{\overline \Psi}^{\, i}_{n-\theta_i}
\Big]
+\frac{1}{2}\,\theta_i^2\Big\}
= N^{(\Psi)} + c^{(\Psi)}~,
\label{L0ferm}
\end{equation}
where again we have distinguished between the number operator
$N^{(\Psi)}$ that counts the fermionic modes and the c-number
$c^{(\Psi)}$ arising from the normal ordering with respect to
$|\Theta\rangle_{\rm NS}$.  In analogy with our discussion of the
bosonic sector, we deduce that this twisted vacuum can be related to
the $\mathrm{Sl}(2,\mathbb{R})$ invariant vacuum $|0\rangle_{\rm NS}$
through the action of $d$ fermionic twist fields $s_{\theta_i}(z)$ of
conformal dimensions
\begin{equation}
h_{s_{\theta_i}}=\frac{1}{2}\,\theta_i^2
\label{hs}
\end{equation}
as follows
\begin{equation}
|\Theta\rangle_{\rm NS} = \lim_{z\to 0}\,\,\prod_{i=1}^d
s_{\theta_i}(z)\,|0\rangle_{\rm NS}~~.
\label{vacuumNS}
\end{equation}
To obtain the conjugate vacuum we use instead fermionic twist fields
$s_{-\theta_i}$ of conformal dimensions
$h_{s_{-\theta_i}}=h_{s_{\theta_i}}$ acting at infinity, namely
\begin{equation}
{}_{\rm NS}\langle-\Theta| = \lim_{z\to \infty}\,{}_{\rm NS}\langle 0|
\prod_{i=1}^d
\Big(s_{-\theta_i}(z)\,z^{2h_{s_{-\theta_i}}}\Big)~~.
\label{bravacuumNS}
\end{equation}
{F}rom (\ref{vacuumNS}) and (\ref{bravacuumNS}) it follows that
\begin{equation}
s_{-\theta_i}(z)\,s_{\theta_j}(w)
\,\sim\,\frac{\delta^{ij}}{(z-w)^{\theta_i^2}}
~~, \quad\quad
\Psi^{\,i}(z)\,{\overline \Psi}^{\,j}(w)\sim
\frac{2\alpha'\,\delta^{ij}}{(z-w)}~,
\label{OPEs}
\end{equation}
which is the fermionic counterpart of (\ref{freeprop}).
Let us now consider some fermionic correlation functions. Using
the mode expansions (\ref{fermexpansions}),
it is easy to prove that
\begin{equation}
{}_{\rm NS}\langle -\Theta|\Psi^{\,i}(z)\,{\overline
\Psi}^{\,j}(w)|\Theta\rangle_{\rm
NS}\,=\,2\alpha'\,\delta^{ij}\left(\frac{w}{z}\right)^{-\theta_i}\frac{1}{(z-w)}
\label{expectation}
\end{equation}
and then deduce for any $i$
\begin{equation}
\frac{\big\langle s_{-\theta_i}(z_1)\,
\Psi^{\,i}(z_2)\,{\overline\Psi}^{\,i}(z_3)
\,s_{\theta_i}(z_4)\big\rangle}{\big\langle s_{-\theta_i}(z_1)
\,s_{\theta_i}(z_4)\big\rangle
\big\langle \Psi^{\,i}(z_2)\,{\overline\Psi}^{\,i}(z_3)\big\rangle}
=
{\omega^{-\theta_i}}~,
\label{fermcorr1}
\end{equation}
where $\omega$ is the anharmonic ratio (\ref{anharmonic}). Other
useful correlators are those involving the first excited states
\begin{equation}
|t_{\theta_i}\rangle_{\rm NS}={\Psi}^{\,\dagger\, i}_{\frac{1}{2}+\theta_i}|\Theta\rangle_{\rm NS}
~~~~{\rm and}~~~~|{\overline t}_{\theta_i}\rangle_{\rm NS}
={\overline \Psi}^{\,\dagger \,i}_{\frac{1}{2}-\theta_i}|\Theta\rangle_{\rm NS}~,
\label{excitedstates}
\end{equation}
whose energy is increased, respectively, of
$(\frac{1}{2}+\theta_i)$ and $(\frac{1}{2}-\theta_i)$ with respect
to the vacuum. Thus, in the $i$-th sector we can introduce excited
fermionic twist fields $t_{\theta_i}(z)$ and
${\overline t}_{\theta_i}(z)$, together with their conjugates $t_{-\theta_i}(z)$ and
${\overline t}_{-\theta_i}(z)$, of conformal dimensions
\begin{equation}
h_{t_{\theta_i}}=h_{t_{-\theta_i}}=\frac{1}{2}(\theta_i+1)^2~~~~{\rm and}~~~~
h_{{\overline t}_{\theta_i}}=h_{{\overline t}_{-\theta_i}}=\frac{1}{2}(\theta_i-1)^2
\label{ht}
\end{equation}
which satisfy
\begin{eqnarray}
{t}_{-\theta_i}(z)\,{t}_{\theta_i}(w)~\sim~
\frac{\delta^{ij}}{(z-w)^{(\theta_i+1)^2}}
~&,&~
{\overline t}_{-\theta_i}(z)\,{\overline t}_{\theta_j}(w)~\sim~
\frac{\delta^{ij}}{(z-w)^{(\theta_i-1)^2}}~~,
\nonumber \\
\frac{{\Psi}^{\,i}(z)}{\sqrt{2\a'}}\,s_{\theta_j}(w) ~\sim~
\frac{\delta^{ij}\,{t}_{\theta_i}(w)}{(z-w)^{-\theta_i}}~&,&~
\frac{{\overline \Psi}^{\,i}(z)}{\sqrt{2\a'}}\,s_{\theta_j}(w) ~\sim~
\frac{\delta^{ij}\,{\overline t}_{\theta_i}(w)}{(z-w)^{\theta_i}}~~.
\label{excitesOPE}
\end{eqnarray}
Proceeding as before, one finds
\begin{subequations}
\begin{align}
&\frac{\big\langle t_{-\theta_i}(z_1)\,
\Psi^{\,i}(z_2)\,{\overline\Psi}^{\,i}(z_3)
\,t_{\theta_i}(z_4)\big\rangle}{\big\langle t_{-\theta_i}(z_1)
\,t_{\theta_i}(z_4)\big\rangle \big\langle
\Psi^{\,i}(z_2)\,{\overline\Psi}^{\,i}(z_3)\big\rangle}
=
{\omega^{-(\theta_i+1)}}~~,
\\
&\frac{\big\langle {\overline t}_{-\theta_i}(z_1)\,
\Psi^{\,i}(z_2)\,{\overline\Psi}^{\,i}(z_3)
\,{\overline t}_{\theta_i}(z_4)\big\rangle}{\big\langle {\overline t}_{-\theta_i}(z_1)
\,{\overline t}_{\theta_i}(z_4)\big\rangle\big\langle
\Psi^{\,i}(z_2)\,{\overline\Psi}^{\,i}(z_3)\big\rangle}
=
\,{\omega^{-(\theta_i-1)}}~~.
\end{align}
\label{fermcorr2}
\end{subequations}

The fermionic correlation function can be alternatively derived thanks
to the bosonization equivalence
\begin{equation}
\label{bosonization}
\Psi^i={\rm e}^{\ii\, H_i}
\quad,\quad   {\cal S}_{\theta^F_i}={\rm e}^{\ii\, \theta^F_i\, H_i}
  \quad,\quad
H_i(z) H_j(w)\sim -\delta_{ij}\,\ln (z-w)~~.
\end{equation}
In this compact notation all previous correlators can be summarized in
\begin{equation}
A^i_{\rm ferm}(z_1,...,z_4|{\cal S}_{\theta^F_i})\equiv
\frac{\big\langle {\cal S}_{-\theta^F_i}(z_1)\,
\Psi^{\,i}(z_2)\,{\overline\Psi}^{\,i}(z_3)
\,{\cal S}_{\theta^F_i}(z_4)\big\rangle}{\big\langle {\cal S}_{-\theta^F_i}(z_1)
\,{\cal S}_{\theta^F_i}(z_4)\big\rangle \big\langle
\Psi^{\,i}(z_2)\,{\overline\Psi}^{\,i}(z_3)\big\rangle}
=
\omega^{-\theta^F_i}~~.
\label{fermcorr4}
\end{equation}
It is interesting to remark that
\begin{equation}
\big\langle
{\cal S}_{-\theta^F_i}(z_1)\,
\Psi^{\,i}(z_2)\,{\overline\Psi}^{\,i}(z_3)
\,{\cal S}_{\theta^F_i}(z_4)
\big\rangle
\,=\,-\,\big\langle
{\cal S}_{-\theta^F_i}(z_1)\,
{\overline\Psi}^{\,i}(z_3)\,\Psi^{\,i}(z_2)
\,{\cal S}_{\theta^F_i}(z_4)
\big\rangle~~,
\label{fermcorr5}
\end{equation}
which is the fermionic counterpart of (\ref{boscorr1}).

This analysis can be generalized to the R sector without any
problems. With R boundary conditions the modes of the fermionic fields
are further shifted with an extra $\frac 12$ with respect to the NS case,
and essentially all occurrences of $\theta_i$ must be
replaced by $\theta_i-\frac 12$. Taking this observation into account,
we can simply
read  the final result from the correlator
(\ref{fermcorr4}) by replacing $\theta_i^F$ with $\theta_i-\ft12$, namely
\begin{equation}
\frac{\big\langle s_{-(\theta_i-\frac 12)}(z_1)\,
\Psi^{\,i}(z_2)\,{\overline\Psi}^{\,i}(z_3)
\,s_{\theta_i-\frac 12}(z_4)\big\rangle}{\big\langle s_{-(\theta_i-\frac 12)}(z_1)
\,s_{\theta_i-\frac 12}(z_4)\big\rangle
\big\langle \Psi^{\,i}(z_2)\,{\overline\Psi}^{\,i}(z_3)\big\rangle
} =\omega^{-(\theta_i-\frac 12)}~~.
\label{fermcorr6}
\end{equation}

\vskip 0.8cm
\sect{Closed String Moduli for Magnetized D-branes}
\label{secn:shifts}

The results of the previous section show that the conformal properties
of open strings moving in a closed string background with a magnetic
field are determined essentially by the shifts $\theta_i$ in the mode
expansions of the various chiral fields.  In this section we will
analyze in more detail how these shifts are related to the closed
string moduli and to the background magnetic field.  To set up the
notation, let us take a $2d$-dimensional torus $\mathcal{T}^{2d}$,
defined in terms of $2d$ real, dimensionless and periodic coordinates
$x^M \sim x^M + 1$, with a constant metric $G_{MN}$ and a constant
anti-symmetric tensor $B_{MN}$, both with dimension of (length)$^2$.
The metric and the $B$ field bring in, respectively, $d(2d+1)$ and
$d(2d-1)$ real parameters, so that our toroidal compactification
depends on a total of $4d^2$ parameters. In
Section~\ref{secn:modeexpansions} we chose to diagonalize the
monodromy matrix $R$ so that $4d^2-d$ parameters are contained by the
vielbein\footnote{The vielbeins $\ECB$ are indeed defined up to
$U(1)^d$ rotations which leave both $\RCB$ and $\GCB$ invariant.}
$\ECB$ and $d$ are encoded in the eigenvalues of $R$. Sometimes it is
convenient to perform a different choice and introduce the vielbein
without making any request on the form of $R$ in the new basis.
This amounts to introducing dimensionful flat coordinates $\tilde \XB$
by $\tilde X^A = E^A_{~M} X^M$ that exhibit no simple periodicity, but
have an orthonormal metric $\delta_{AB}$. If we impose no further
conditions, the anti-symmetric background $\tilde \BB$ (dimensionless)
remains generic in such a frame, and the choice of frame is ambiguous
up to $\mathrm{SO}(2d)$ rotations. Thus the vielbein $\EB$ contains
$(2d)^2 - d(2d-1)$ independent parameters, just as the metric $\GB$.
We can split the orthonormal coordinate in two groups: $\tilde X^A \to
(\tilde X^a,\tilde Y^a)$. We introduce then complex coordinates $Z^a =
(\tilde X^a + \ii \tilde Y^a)/\sqrt{2}$, {\it i.e.} in matrix notation we
set
\begin{equation}
\ZB = \left(\begin{matrix}
Z\cr {\bar Z}\end{matrix}\right) =
\SB \left(\begin{matrix}\tilde X\cr  \tilde
Y \end{matrix}\right)~~~~,~~~~
\SB = \frac{1}{\sqrt{2}} \left(\begin{matrix} 1 & \ii\cr 1 &
    -\ii\end{matrix} \right)~~.
\end{equation}
The parameters of the metric $G$ are now encoded in the complex
vielbein $\EB'= \SB\EB$, which is defined up to the
realization of $\mathrm{SO}(2d)$ over the complex frame $\ZB = (Z,
\bar Z)$. In presence of an antisymmetric tensor, we can partially fix
this ambiguity by requiring that, in the complex frame, it is of type
$(1,1)$. For instance in the heterotic context, it is natural to use
the $B$-field and fix the vielbein $E'$ so that
\begin{equation}
\label{B11}
\BB' = {}^t(\EB')^{-1} \BB (\EB')^{-1} = \left(\begin{matrix} 0 & \ii
    b \cr - \ii \bar b & 0\end{matrix}\right)~~~~,~~~~
(b^\dagger = b)~~.
\end{equation}
This form is invariant under the $\mathrm{U}(d)$ subgroup of
$\mathrm{SO}(2d)$ acting block-diagonally on the complex frame: $Z \to
U Z$, $\bar Z \to \bar U \bar Z$ and can be imposed by means of
$\mathrm{SO}(2d)/\mathrm{U}(d)$ transformations.  The residual
$\mathrm{U}(d)$ invariance can be used, for instance, to put the
complex vielbein $\EB'$ in the form
\begin{equation}
\label{E'choice}
\EB' = \frac{1}{\sqrt{2}} \left(\begin{matrix}V & 0 \cr 0 &
   V\end{matrix}\right) \left(
\begin{matrix}1 & U \cr 1 & \bar U\end{matrix}
\right)
\end{equation}
with $V$ real.

In Type I theories, it is more natural to ask the property~\eq{B11}
for ${\cal F}_\sigma$.  It might be impossible to choose a complex
structure so that all ${\cal F}_\sigma$'s are $(1,1)$ forms, in which
case supersymmetry is broken~\cite{Marino:1999af}. We will return on
this point in Section~\ref{secn:kahler}, when we compute the v.e.v.
of the auxiliary fields $D$ and $F$. Now let us just notice that, if
all ${\cal F}_\sigma$'s are $(1,1)$ forms, the reflection matrices
$R'_\sigma$ are block diagonal in the complex basis
\begin{equation}
R'_\sigma =
\begin{pmatrix}
 r'_\sigma & 0 \\ 0 & \overline r'_\sigma
\end{pmatrix}~~.
\end{equation}

There are several different ways to organize the compactification
moduli which we denote generically by $m$.  When we are interested in
holomorphicity properties, then it is convenient to use the elements
of the matrix $U$ in Eq.~\eq{E'choice}, which are directly related to
the complex structure. In fact, the mixed tensor $ \ii dz^i\otimes
\partial_{z^i} -\ii d{\overline z}^i\otimes \partial_{{\overline
    z}^i}$ depends only on $U$, when written in the (original) real
basis (in the Type I case, the K\"ahler structure arises from the
complexification of $V$~\eq{E'choice} with the R-R $2$-form).
Alternatively, when one deals with non-holomorphic terms, it is more
natural to associate the moduli $m$ directly to the $G$. To write the
vertex operator associated to a generic modulus, let us recall that
the closed string coordinates are given by
\begin{equation}
x^\mu(z,\overline z) = \frac{1}{2}\Big[ X^\mu_L(z) + X^\mu_R(\overline z)\Big]
~~~~{\rm and}~~~~
x^M(z,\overline z) =\frac{1}{2}\Big[ X^M_L(z) + X^M_R(\overline z)\Big]
\label{xclosed}
\end{equation}
where the index $\mu$ labels the uncompact directions,
and the subscripts $L$ and $R$ denote, respectively, the left
and right moving parts. For example one has, for any $z \in \mathbb{C}$,
\begin{equation}
X^M_L(z) = q^M_L-\ii 2\a'\,p^M_L\,\log z + \ii\sqrt{2\a'}\sum_{n=1}^\infty\Big[
\frac{{a}_{L\, n}^{~M}}{n}\,z^{-n}
-\frac{a^{\dagger\,M}_{L\, n}}{n}\,z^{n}\Big]~~.
\label{xl}
\end{equation}
The fermionic string coordinates admit a similar left/right
decomposition.
The mode expansion for $X^\mu$ and $\psi^\mu$ is formally identical to that
of $X^M$ and $\psi^M$, the only difference being that the former are chosen to be dimensionful.
The massless closed string excitations of the NS-NS
sector that represent fluctuations of the metric $G$ or of the $B$
field along the compact directions\footnote{As already mentioned, in a
  complete orientifold compactification the $B$ field is not
  dynamical; however we will formally consider it on the same footing
  as $G$, in order to derive the dependence of the effective action on
  the possible discrete values $B$ can have~\cite{Bianchi:1991eu}.}
are described by the usual vertex operators $V^M_L(z)\,V^N_R(\overline
z)$, where (in the 0-superghost picture) we have
\begin{subequations}
\label{vLR}
\begin{align}
&V^M_L(z) = \Big[\partial X^M_L(z) + \ii
(k_L\cdot\Psi_L)\Psi_L^M(z) \Big]\,{\rm e}^{\ii\,k_L\cdot X_L(z)}~~,
\\
&V^N_R(\overline z) = \Big[\bar\partial X^N_R(\overline z) + \ii
(k_R\cdot\Psi_R)\Psi_L^N(\overline z) \Big]\,{\rm e}^{\ii\,k_R\cdot X_R(\overline
z)}~~.
\end{align}
\end{subequations}
In these expressions $k_L$ and $k_R$ denote the left and right momenta
of the emitted state while the symbol $\cdot$ is a shortcut for the vector product
with metric $\eta_{\mu\nu}$.
In general, when $k_L\not=k_R$, a more careful
definition of the vertex operator is necessary to ensure the bosonic
character of the operators $V_{L,R}$~\cite{Polchinski:1998rq}
\begin{equation}
W^{MN}(z,\overline{z}) = \ex{-\ii\pi\a' (k_L+k_R)\cdot p_L} V^M_L(z)\;
\ex{ \ii\pi\a' (k_L+k_R) \cdot p_R} V^N_R(\overline z)~~.
\label{vMN}
\end{equation}
Actually we are not interested in Kaluza-Klein modes and we take $k_L$
and $k_R$ to be aligned entirely along the uncompact directions.
However, we perform a slight off-shell extension of the closed string
vertices, by formally taking, along the uncompact directions,
$k_L^2=k_R^2=0$ with $k_L\not= k_R$. In this way we can have
$(k_L+k_R)^2 \not=0$, without spoiling the conformal properties of the
left (or right)  part of the vertex.

The insertion of the operator (\ref{vMN}) inside a string correlation
function induces a variation of $G_{MN}$ and of $-B_{MN}$, associated
respectively to the symmetric and anti-symmetric parts in the indices
$M$ and $N$~\footnote{This can be seen by using~\eq{Sx} and taking the
  $\partial_m$ derivative of the Euclidean weight ${\rm e}^{-S}$
  present in the path integral; notice that the normalization
  of~\eq{vm} depends also on our convention~\eq{xclosed}.}.
Thus, the variation due to a change in a
generic modulus $m$ is produced by the following vertex operator
\begin{equation}
W_m(z,\overline z) =
\frac{1} {4\pi\alpha'}\frac{\partial}{\partial
m}(G-B)_{MN} W^{MN}(z,\overline z)~~.
\label{vm}
\end{equation}
Since in
toroidal compactifications the vertex (\ref{vm}) represents a truly
marginal deformation for any value of $m$, we can schematically  write
\begin{equation}
\int \!d^2 z\,\langle \cdots W_m(z,\overline z) \cdots \rangle
= \frac{\partial}{\partial m}
\langle \cdots \rangle
\label{amplvar}
\end{equation}
where $\cdots$ stand for any sequence of string vertex operators.
The partial derivative with respect to $m$ is taken by keeping fixed at
arbitrary values all other moduli, which are indeed described by
independent vertices. As it is intuitively natural, two different vertices
$W_m$ and $W_{m'}$ are independent if the corresponding states are
orthogonal, {\it i.e.} $\langle m'| m\rangle=0$. For instance, the four
dimensional dilaton $\phi_4$ is clearly independent of the moduli~\eq{vm}
specifying the compact space, since it involves only string coordinates
along the Minkowski directions. This means that the differential equation
we derive from~\eq{amplvar} are computed by keeping $\phi_4$ fixed. As it
shown in~\cite{Lust:2004cx}, the dependence on the four dimensional
dilaton can be derived in the same way by inserting in~\eq{amplvar} the
appropriate vertex $W_{\phi_4}$.

Let us now introduce a stack of D-branes wrapped on $\mathcal{T}^{2d}$.
On their world-volume we may introduce a background field $\FB$
whose components $F_{MN}$ are quantized as
\begin{equation}
\label{qc}
\frac{1}{2\pi} F_{MN}= \frac{p_{MN}}{l_M l_N}
\end{equation}
where $p_{MN}$ is the standard Chern class and $l_M$ is the
wrapping number of the D-brane around the cycle $dX^M$.
As discussed in Section \ref{secn:modeexpansions}, the open strings
connecting two such D-branes are described in terms of twisted bosonic
and fermionic fields, whose monodromy matrix $\RB = \RB_\pi^{-1}\RB_0$
is defined in terms of the boundary reflection matrices $\RB_\sigma$
given in (\ref{R}).  These conformal fields and their correlation functions
are described in the orthonormal complex basis introduced
in Eqs. (\ref{diagonalR1}) and (\ref{Gcal}), so that all relevant
information is encoded  entirely in the $d$ phases $\theta_i$. Of course
these twists, as well as the complex vielbein $\ECB$, depend on the $4d^2$ parameters
contained in $G$ and $B$.  In the next sections we will compute mixed
amplitudes with insertions of closed string vertex operators $V_m$
inside correlators of twisted open strings, which, as indicated in
(\ref{amplvar}), account for the derivatives with respect to a NS-NS
modulus $m$. For the physical interpretation of the results it will be
crucial to know how the twists $\theta_i$ depend on $m$. In particular
it will be important to know the derivatives of $\theta_i$ with
respect to $m$. As shown in detail in Appendix \ref{app:jac}, these
are given by
\begin{equation}
\label{dmt8}
\begin{aligned}
2\pi\ii\, \frac{\partial\theta_i}{\partial m}
=\left( \frac{\partial {\cal R} }{\partial m} \, {\cal R}^{-1}\right)_{ii} ~=~ &
\frac 12\, \left(\ECB\,
 \GB^{-1} \,\frac{\partial(\GB - \BB)}{\partial m}\, \left[\RB_\pi - \RB_0\right]
\,\ECB^{-1}\right)_{ii}
\\
& -   \,\frac 12\, \left(\ECB
\left[\RB_\pi^{-1} - \RB_0^{-1}\right]\,\GB^{-1} \,\frac{\partial(\GB + \BB)}{\partial
m}
\,\ECB^{-1}\right)_{ii}~~.
\end{aligned}
\end{equation}
It is worth noticing the appearance in this formula of the same
expression that plays the role of the polarization in the vertex
operator (\ref{vm}). Eq. (\ref{dmt8}) applies to a generic toroidal
configuration with any value of $\GB$ and $\BB$, and to generic ({\it
  i.e.} non-commuting) magnetic fluxes $\FB_\sigma$ on the wrapped
D-branes. To make contact with the set-up that is usually considered
in the literature, and as an illustration, we now consider the simple
case of D-branes on factorized torii with diagonal fluxes, which are
T-dual to a system of intersecting D-branes at angles.

\vskip 0.5cm
\subsection{Factorized torus with commuting fluxes}
\label{subsecn:fac_diag}
Let us consider a model in which the internal torus $\mathcal{T}^6$ is
metrically factorized as
$\mathcal{T}^2_{(1)}\times\mathcal{T}^2_{(2)}\times\mathcal{T}^2_{(3)}$.
Let us also assume that the background NS-NS field $\BB$ and the gauge
fields $\FB_\sigma$ respect this factorized structure.  In this case
we can treat each torus $\mathcal{T}^2_{(i)}$ independently of the
others so that the problem becomes two-dimensional and drastically
simplifies.  In each torus the real metric and $B$-field can be
parameterized in terms of two complex moduli $T= T_1 + \ii T_2$ and
$U= U_1 + \ii U_2$ as follows
\begin{equation}
\label{GB2}
\GB = \alpha' \,\frac{T_2}{U_2}\,
\begin{pmatrix}1 & U_1 \\ U_1 & |U|^2\end{pmatrix}~~~~{\rm and}~~~~
\BB = \alpha'\,
\begin{pmatrix}\,
0 & - T_1 \\ T_1 & 0
\end{pmatrix}~~.
\end{equation}
This parameterization with $T$ and $U$ is very convenient
to discuss the effects of simple T-duality transformations. Indeed,
a T-duality along the $x=x^1$ axis
amounts just to the exchange $T \leftrightarrow U$, while a
T-duality along $y=x^2$ corresponds to $T \leftrightarrow -1/U$.
On each torus the magnetic fluxes are of the form
\begin{equation}
\label{F2d}
2\pi\alpha' \FB_\sigma = \alpha'
\begin{pmatrix}
0 & f_\sigma \\ - f_\sigma & 0
\end{pmatrix}~~,
\end{equation}
where $f_\sigma$ is real and quantized according to
Eq.~(\ref{qc}).
We can use the complex vielbein
\begin{equation}
\label{E2d}
\ECB = \sqrt{\frac{\alpha' T_2}{2 U_2}}
\begin{pmatrix}
1 & U \\ 1 & \overline U
\end{pmatrix}~~~~{\rm and}~~~~
\ECB^{-1} = \ii \sqrt{\frac{\alpha'}{2 T_2 U_2}}
\begin{pmatrix}
\overline U & - U \\ -1 & 1
\end{pmatrix}
\end{equation}
to put the metric in the form (\ref{Gcal}). In the resulting complex basis $\ZCB
= \ECB X$ it is straightforward to compute the reflection matrices
$\RCB_\sigma = \ECB \RB_\sigma \ECB^{-1}$ by specializing their
definition (\ref{R}) to the present case and using \Eq{E2d}. The
result is
\begin{equation}
\label{Rs2d}
\RCB_\sigma = - \diag
\left(
\frac{\overline T - f_\sigma}{T - f_\sigma}, \frac{T - f_\sigma}{\overline T - f_\sigma}
\right)~~.
\end{equation}
In this basis the monodromy matrix is diagonal $\RCB' = \diag
\left(\ex{2\pi\ii\theta},\ex{-2\pi\ii\theta}\right)$ with
\begin{equation}
\label{theta2d}
\ex{2\pi\ii\theta} = \frac{T - f_\pi}{\overline T - f_\pi}
\,\, \frac{\overline T - f_0}{T - f_0}~~.
\end{equation}
These formulas will be useful in later sections to make contact
with some existing results in the literature.

\sect{Low-Energy Spectrum on D-branes with Fluxes}
\label{secn:spectrum}

In this section we recall the main features of the open string
low-energy spectrum for systems of D9-branes with general magnetic
fluxes. In a system with two or more stacks of D9-branes, there are
two classes of open strings: those that start and end on the same set
of D9-branes, and those which connect D9-branes with different
magnetic fields. The first type of open strings give rise to
``untwisted'' states transforming in the adjoint representation of the
gauge group living on the D9's under consideration. In what follows we
will focus on the second type of open strings related to ``twisted''
states transforming in the bi-fundamental representation.  In the NS
sector, the complete Hamiltonian for this twisted open string is
\begin{equation}
\label{ham}
H^{\rm NS}  = L_0^{\rm x\,\psi} + L^{({\cal Z})}_0 + L_0^{(\Psi)} -
\frac 12~,~~\mbox{with}~~~
 L_0^{\rm x\,\psi} = \Big(\alpha' p_\mu\, p^\mu + \sum_{n=1} a_n^{\dagger\,\mu}
a_n^\mu + \sum_{r=\frac 12} r ~\psi_{r}^{\dagger\,\mu}
\psi_{r}^{\mu} \Big)~~.
\end{equation}
By using Eqs.~(\ref{L0}) and~(\ref{L0ferm}) for $L^{({\cal Z})}_0$ and
$L_0^{(\Psi)}$ , we can express the mass-shell condition for the NS
sector as follows
\begin{equation}
\Big(L_0^{\rm x\,\psi} + N^{(\mathcal{Z})} + N^{(\Psi)} - \frac 12
  + \frac{1}{2} \sum_{i=1}^3 \theta_i\Big)\,|\phi\rangle_{\rm NS} \,=\,0~~.
\label{massshell}
\end{equation}
Finally, in order to define the physical spectrum, we should specify
the GSO projection. In the NS sector, the GSO projection on open
strings stretched between two D-branes is defined to remove the vacuum
and to select only those states with an odd number of fermionic
oscillators acting on it. The opposite choice would describe an open
string stretched between a D-brane and an anti-D-brane. It follows
from the observation made just after Eq.~\eq{fermvacuum} that we can
now interpolate continuously between these two situations.
In fact, when one of the angles $\theta_i$ is bigger than $1/2$ the
usual GSO projection with respect to $|\Theta\rangle_{\rm NS}$ selects
the vacuum ({\it i.e.}  ${\overline\Psi}_{\frac
  12-\theta_i}^{\,\dagger \,i} |\Theta\rangle_{\rm NS}$) as well as
all states with an even number of fermionic oscillators acting on
it\footnote{The case $\theta_i=\frac 12$ is special and requires a
  separate treatment due to the appearance of zero modes.}. Thus
we have two possibilities: we can limit the
range of the angles to $[0,1/2]$ and specify in each case whether we
take the brane/brane or the brane/anti-brane GSO; otherwise we keep
the interval $0\leq\theta_i<1$, but we stick always to the same GSO.
Here we will use this second option. Then the first low-lying states
in the spectrum are:
\begin{subequations}
\label{statesNS}
\begin{align}
&{\rm 1~vector}~~~~~~\psi_{\frac{1}{2}}^{\dagger\,\mu}\ket{k;\Theta}_{\rm NS}
& 2\alpha'M^2 = & \sum_{j=1}^3 \theta_j~,
\label{vector}\\
&{\rm 3~scalars}~~~~~~\Psi_{\frac{1}{2}+\theta_i}^{\,\dagger \,i}\ket{k;\Theta}_{\rm NS}
&  2\alpha'M^2_i = & \sum_{j\not=i}^3 \theta_j + 3\theta_i~,
\label{scalars1}\\
&{\rm 3~scalars}~~~~~~{\overline \Psi}_{\frac{1}{2}-\theta_i}^{\,\dagger\, i}
\ket{k;\Theta}_{\rm NS}
& 2\alpha'M^2_i = & \sum_{j\not=i}^3 \theta_j - \theta_i~.
\label{scalars2}
\end{align}
\end{subequations}
where $\ket{k;\Theta}_{\rm NS}$ is the twisted vacuum with
four dimensional momentum $k^\mu$.  With our convention, it is clear
that the vector (\ref{vector}) and the three scalars (\ref{scalars1})
never contribute to the low-energy spectrum except when all
$\theta_i$'s are small. In fact they can survive the field theory
limit $\alpha'\to 0$, only if all $\theta_i$ goes to zero as $\a'$
does. On the contrary, some of the scalars (\ref{scalars2}) may remain
in the effective theory also for non-zero twists: for particular values
of the $\theta_i$'s they are massless, but in general they are massive.
To appreciate better this point, let us write the twists as
follows\footnote{We recall that a behavior like (\ref{thetaalpha}) is typical in the
instanton sector of non-commutative gauge theories realized with open strings in a
non-trivial $B$ background. In fact, some of the instanton moduli
correspond to twisted open strings for which the sub-leading
corrections $\epsilon_i$ are related to the dimensionful non-commutativity
parameter~\cite{Billo:2005fg}. Moreover, a scaling behavior like (\ref{thetaalpha}) has
been considered also in the field theory analysis of intersecting
brane models~\cite{Cremades:2004wa}.}
\begin{equation}
\theta_i = \theta_i^{(0)} + 2\alpha' \epsilon_i
\label{thetaalpha}
\end{equation}
where $\theta_i^{(0)}$ and $\epsilon_i$ are quantities which are
kept fixed in the limit $\alpha'\to 0$. In other words,
$\theta_i^{(0)}$ is the ``field theory'' value of the $i$-th twist, while
$\epsilon_i$, which has dimensions of a (mass)$^2$,
is its sub-leading string correction.
Inserting (\ref{thetaalpha}) in the mass formula (\ref{massshell}), we
find
\begin{equation}
M^2_i = \frac{1}{2\alpha'}\left(\sum_{j\not=i}^3 \theta_j^{(0)} -
\theta_i^{(0)}\right)
+ \sum_{j\not=i}^3 \epsilon_j - \epsilon_i~~.
\label{mass2}
\end{equation}
Therefore, by suitably choosing the $\theta_i^{(0)}$'s we
can cancel the term in brackets and obtain, in the limit $\alpha'\to
0$, a {\it finite} mass for some of the states
(\ref{scalars2})\footnote{Using (\ref{thetaalpha}) in the mass
  formulas (\ref{vector}) and (\ref{scalars1}), we may find a finite
  non-zero mass for the vector (\ref{vector}) and the scalars
  (\ref{scalars1}) only if all $\theta_i^{(0)}$'s are zero.}.  Thus,
the spectrum is in general non-supersymmetric, but it is known
that the presence of a non-trivial mass~\eq{mass2} breaks
supersymmetry spontaneously. In the field theory limit this
breaking appears simply as a Fayet-Iliopoulos term due to the
presence of non-trivial v.e.v.'s of the auxiliary field $D$ in the
$U(1)$ gauge
superfields~\cite{Cvetic:2001tj,Cvetic:2001nr,Cremades:2002te}.
This observation will be important for our future calculations and
we will give a direct stringy proof of this statement in
Section~\ref{secn:kahler}.

In view of these considerations, from now on we will focus on the
scalars (\ref{scalars2}), which we denote by $\phi^i$.  Recalling our
discussion of Section \ref{secn:modeexpansions} and adopting the
notation presented there, we can see that the vertex operator for the
emission of $\phi^i$ with momentum $k^\mu$ is (in the
$(-1)$-superghost picture)
\begin{equation}
V_{\phi^i}(z) = \phi^i(k) \,
\prod_{j=1}^3 \left( {\cal S}_{\theta_{j\,(i)}^{F}}\!(z) \,
\sigma_{\theta_j}(z)\right)
{\rm e}^{- \varphi(z)} \,{\rm e}^{\ii \,k \cdot X(z)}
\label{vertexphi}
\end{equation}
where $\varphi$ is the chiral boson of the superghost bosonization
formulae, and $\sigma_{\theta_j}$ and ${\cal S}_{\theta_{j\,(i)}^{F}}$ are
the bosonic and fermionic twist fields. The labels of the latter
are
\begin{equation}
{\theta_{j\,(i)}^{F}}= \left\{
 \begin{array}{lll}
 \theta_j&{\rm for}&j\not = i\\
 \theta_j-1&{\rm for}&j=i
 \end{array}
\right.
\end{equation}
which, according to Eq.~(\ref{bosonization}), correspond to take
\begin{equation}
{\cal S}_{\theta_{j\,(i)}^{F}}\!(z) =
\left\{
 \begin{array}{lll}
s_{\theta_j}(z)&{\rm for}&j\not = i~~\\
{\overline t}_{\theta_j}(z)&{\rm for}&j=i~~.
\end{array}
\right.
\label{sthetaFj}
\end{equation}
One can easily check that the vertex (\ref{vertexphi}) has conformal
dimension 1 if the mass-shell condition (\ref{scalars2})
is satisfied.

The complex conjugate scalars ${\overline \phi}^{\,i}$ are
associated to twisted open strings with the opposite orientation
as compared to those considered so far, and thus their
corresponding vertex operators (again in the $(-1)$-superghost picture) are
\begin{equation}
V_{{\overline \phi}^{\,i}}(z) = {\overline \phi}^{\,i}(k) \,
\prod_{j=1}^3 \left( {\cal S}_{-\theta_{j\,(i)}^{F}}\!(z) \,
\sigma_{-\theta_j}(z)\right)
{\rm e}^{- \varphi(z)} \,{\rm e}^{\ii \,k \cdot X(z)}~~.
\label{vertexbarphi}
\end{equation}
Finally, we remark that
the polarizations $\phi^i$ and ${\overline \phi}^{\,i}$ of the
vertices (\ref{vertexphi}) and (\ref{vertexbarphi}) contain the appropriate Chan-Paton factors
for the bi-fundamental representations of the gauge group,
and have dimensions of (length)$^{-1}$ in units of $2\alpha'$.

Let us now consider the twisted R sector. For generic values of the twists
$\theta_i$'s, only the four fermionic coordinates $\psi^\mu$ along the
uncompact directions have zero modes, and thus the vacuum will carry a
spinor representation of the four-dimensional Lorentz group
$\mathrm{SO}(1,3)$. Furthermore, in the R sector the GSO projection
selects a definite chirality (say positive) for such a spinor, which
therefore can be denoted by $\ket{\alpha,k;\Theta}_{\rm R}$, with
$\alpha$ being a chiral spinor index.  The complete Hamiltonian
$H_0^{\rm R}$ of the R sector is given by the obvious generalization
of (\ref{ham}) in which the $\psi^\mu$'s have integer moding and the
twisted fermions are as in (\ref{fermexpansions}) with $\nu=0$. As a
consequence, there is a cancellation between the bosonic and fermionic
c-number terms due to normal ordering, so that $c^{\rm R}=0$, and
the mass-shell condition for any state $\ket{\phi}_{\rm R}$ is
\begin{equation}
L_0^{\rm R}\,\ket{\phi}_{\rm R} = 0~~.
\label{masssshellR}
\end{equation}
Applying this formula to the vacuum $\ket{\alpha,k;\Theta}_{\rm R}$, we can deduce
that $k^2=0$ for any non-zero value of the twists $\theta_i$. The vertex operator
associated to such a massless spinor, which we will denote by $\lambda_\alpha$,
is (in the $(-1/2)$-superghost picture)
\begin{equation}
V_{\lambda}(z) = \lambda_\alpha(k)\,S^{\alpha}(z) \prod_{j=1}^3
\left( s_{\theta_j-\frac 12}(z)\,
\sigma_{\theta_j}(z)\right)
{\rm e}^{- \frac 12\,\varphi(z)} \,{\rm e}^{\ii \,k \cdot X(z)}
\label{vertexR}
\end{equation}
where $S^\alpha$ is the chiral spin-field of $\mathrm{SO}(1,3)$
and the polarization $\lambda_\alpha$ has dimensions of
(length)$^{-3/2}$ in units of $2\alpha'$.
One can easily check that this vertex operator has conformal
dimension 1 if $k^2=0$.

When one of the twist parameters is zero, one of the internal complex
fermions $\Psi^i$ ceases to be twisted and two extra fermionic real
zero-modes appear. In this case the vacuum becomes doubly degenerate
and one finds two massless fermions in four dimensions. When all
twists are vanishing, all internal fermions have zero-modes and, upon
compactification, one finds four massless fermions in the resulting
four-dimensional theory.

\begin{figure}[ht]
\begin{center}
\includegraphics[width=0.35\textwidth]{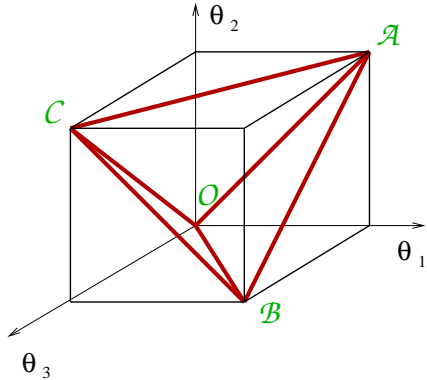}
\caption{\label{tetrahedron}\small The tetrahedron in $\theta$-space.}
\end{center}
\end{figure}

In summary, the low-energy spectrum of open strings stretched
between two stacks of D9 branes consists of one chiral massless
fermion and a number of scalars that are generically massive (or
tachyonic). For specific values of the fluxes and hence of the
twists, one or more scalars may become massless and supersymmetric
configurations may be realized. This situation can be conveniently
represented in terms of a tetrahedron in the twist parameters
space \cite{Rabadan:2001mt}, as shown in Fig. \ref{tetrahedron}.
This represents supersymmetric configurations and separate an
inner region, where the scalars are all massive, from an outer
region, where the scalars become tachyonic. Faces, edges and
vertices of the tetrahedron correspond to $\mathcal{N}=1$,
$\mathcal{N}=2$ and $\mathcal{N}=4$ configurations, respectively.
Notice that in our conventions, where the twists $\theta_i$'s are
taken in the range $[0,1)$, the vertices $\mathcal{A}$,
$\mathcal{B}$ and $\mathcal{C}$, and the face $(\mathcal{ABC})$
are in fact not part of the moduli space. Of course one could change
conventions and choose a different parameterization without changing
the physical conclusions. We will briefly return on
this point in Section~\ref{secn:kahler}.

\vskip 0.8cm
\sect{Moduli Dependence of the K\"ahler Metric}
\label{secn:kahler}

In this section we compute the dependence on the closed string moduli of the
K\"ahler metric for the chiral matter in the effective action
of magnetized D9-branes. Exploiting T-duality, this system can be used also
for brane-worlds involving branes at angles with
arbitrary open string fluxes. Our analysis generalizes previous
results in the literature since we obtain an expression for the
K\"ahler metric that is valid not only for commuting
and supersymmetric fluxes on factorized torii, but also for arbitrary
{\it non-commuting} and {\it non-supersymmetric} configurations on
{\it generic} torii.

Let us then consider the chiral fields (\ref{scalars2}) arising from
$\theta$-twisted open strings connecting two (stacks of) D9-branes
with generic non-commuting open string fluxes. The moduli dependence
of the K\"ahler metric can be extracted from a 3-point function on a
disk involving two open twisted matter fields $\phi^i$ and $\bar
\phi^{i}$, and one closed string parameter $m$, namely
\begin{equation}
\label{ampdef}
{\cal A}_{m \bar\phi^i \phi^i} =\frac{1}{2\alpha'}\,  \langle
V_{\bar\phi^{i}}\, W_m \,V_{\phi^i}\rangle ~~.
\end{equation}
The normalization in this amplitude can be obtained by using the
results of Appendix~A of Ref.~\cite{DiVecchia:1996uq}\footnote{In
  particular it sufficient to use Eq.~(A.16) of that paper with the caveat
  that the normalization of the twisted scalar (\ref{vertexphi}) is
  one-half of the normalization of the gluon vertex operator.},
but, as we shall see,
it can also be fixed independently by using the value of the tree-level mass given in
(\ref{scalars2}).

In order to extract the K\"ahler metric
from  ${\cal A}_{m \bar\phi^i \phi^i}$,
a few steps must be performed. First of all, to write
the results in terms of (properly normalized) field theory quantities,
the open string vertices should be transformed from the canonically normalized
string theory basis to the field theory basis according to
\begin{equation}
V_{\phi^i} \rightarrow \left(K_{ii}\right)^{1/2}
V_{\phi^i}~~~~,~~~~
V_{\bar \phi^i} \rightarrow \left(K_{i i}\right)^{1/2}
V_{\bar \phi^i}
\label{rescaling}
\end{equation}
where $K_{ii}$ is the K\"ahler metric for the $i$-th
matter multiplet. Second, a factor of $\ii$ must be introduced to transform the string scattering amplitude
into the corresponding term in the effective action.
The string/field theory dictionary then reads
\begin{equation}
\label{kahmod}
{\cal A}_{m\bar \phi^i \phi^i }  = {\ii}\,
\left(K_{ii}\right)^{-1}
~\frac{\partial}{\partial m}
~{\cal L}^{(2)}~~,
\end{equation}
where ${\cal L}^{(2)}$ is the quadratic part of Lagrangian of the twisted scalars
with the field theory normalization.
In our conventions, with a mostly plus metric, this Lagrangian reads
\begin{equation}
{\cal L}^{(2)} = -\,K_{ii}
\Big(\partial\bar
\phi^i \cdot \partial \phi^i +M_i^2\,\bar \phi^i\,\phi^i\Big)~~.
\label{lagrangians1}
\end{equation}
Thus, from (\ref{kahmod}) and going to
momentum space, one finds
\begin{equation}
\begin{aligned}
{\cal A}_{m\bar \phi^i \phi^i } &= {\ii}\,
\left(K_{ii}\right)^{-1}
\,\frac{\partial}{\partial m} \Big[
\left(
\,  k_1\cdot k_2 -
\,M_i^2  \right)K_{ii}\Big]\,\overline\phi^i(k_1)\,\phi^i(k_2)\\
&=\ii\Big[-\frac{\partial M^2_i}{\partial m}
+\big(k_1\cdot k_2 - M^2_i\big)\,\frac{\partial\ln K_{ii}}{\partial m}
\Big]\,\overline\phi^i(k_1)\,\phi^i(k_2)~~,
\end{aligned}
\label{lagder}
\end{equation}
where in the second line we have explicitly taken into account the
dependence of the mass $M^2_i$ on the open string twists
and hence on the closed string moduli.

The correlator (\ref{ampdef}) is a mixed open/closed string
amplitude which can be computed after writing the closed string
vertex operator $W_m$ in terms of the propagating (twisted) open string. This
is done by using the boundary conditions
on the disk discussed in Section \ref{secn:modeexpansions}, which imply
in particular
\begin{equation}
V_L^M(z)=V^M(z)~~~~,~~~~V_R^M(\bar z)=(R_0)^M_{~N}\,V^N(\bar z)
\label{openclosed}
\end{equation}
along the compact directions\footnote{Notice that since the vertex operators
$V^M$ are written in terms of twisted conformal fields, the reflection rules
(\ref{openclosed}) automatically take into account the presence of
different boundary conditions on the disk.}, and
\begin{equation}
V_L^\mu(z)=V^\mu(z)~~~~,~~~~V_R^\mu(\bar z)=V^\mu(\bar z)
\label{openclosed1}
\end{equation}
along the uncompact ones.
Thus, the amplitude (\ref{ampdef}), including the cocycle
introduced in Eq. (\ref{vMN}), can be written as
\begin{equation}
\label{corr4}
{\cal A}_{m \bar \phi^i \phi^i} =
\frac{{\rm e}^{-\ii\pi \alpha' k_L \cdot k_R}}{8\pi {\alpha'}^2} \,
\left[ \frac{\partial}{\partial m} ( G- B)\cdot R_0\right]_{MN}\,
\langle V_{\bar \phi^i}\, V^M V^N \,V_{\phi^i}\rangle~~.
\end{equation}
In the orthonormal basis (\ref{diagonalR1}), one finds
\begin{equation}
{\cal A}_{m\bar \phi^i \phi^i} =
\left[ \frac{\partial}{\partial m} ( G- B)\cdot R_0\right]_{MN}
\,({\bf \cal E}^{-1})^M_{~a} \, ({\bf \cal E}^{-1})^N_{~b} ~ {\cal A}_{(i)}^{ab}
\label{deriv0}
\end{equation}
where the indices $a,b=(i,\bar{i})=1,...6$ span the orthonormal
frame, in which the monodromy matrix $R$ is diagonal and the metric $G$ is of the form
(\ref{Gcal}), and $({\cal E}^{-1})^M_{~a}$ is the inverse of the vielbein ${\cal
E}^a_{~M}$ introduced in Eq.~(\ref{diagonalR1}). The matrix ${\cal
A}_{(i)}$ is explicitly given by
\begin{equation}
{\cal A}_{(i)} \equiv
\left(
\begin{array}{cc}
  0 & A_{j(i)}\, \delta^{j{\overline k}} \\
  {A}_{{\overline j} (i)}\, \delta^{{\overline j}k} & 0 \\
\end{array}
\right)
\label{calAij}
\end{equation}
where
\begin{equation}
{A}_{j(i)}
= \frac{{\rm e}^{-\ii\pi\alpha' k_L \cdot k_R}}{8\pi {\alpha'}^2}\,
 \langle V_{\bar \phi^i}\, V^j  V^{\overline j} \,
V_{\phi^i}\rangle~~,
\label{4ptbasic}
\end{equation}
and ${A}_{{\overline j} (i)}$ is given by the same expression
(\ref{4ptbasic}) with $V^j$ and $V^{\overline j}$ exchanged.

The string correlator ${A}_{j(i)}$ has to be computed in the orthonormal basis, where one
can use the CFT results summarized in
Section~\ref{secn:modeexpansions}.
It is important to notice that although ${A}_{j(i)}$ depends
only on the open string twists $\theta_i$'s, the full amplitude ${\cal A}_{m\bar \phi^i \phi^i}$
contains additional dependencies on the various closed string moduli through
the reflection matrix $R_0$ and the inverse vielbein ${\cal E}^{-1}$.
In summary,
the computation of the string amplitude ${\cal A}_{m\bar \phi^i \phi^i}$
elegantly separates into two pieces: the correlator $A_{j(i)}$ and a prefactor
that carries the information on the specific closed string modulus inserted
and the boundary conditions.
Let us start computing the first piece.

\vskip 0.5cm
\subsection{The string correlator}

Here we derive the four point function  ${\cal A}_{j (i)}$ defined
in (\ref{4ptbasic}).
For the chiral matter fields $\phi^i$ and $\overline \phi^i$
we take the vertex operators (\ref{vertexphi}) and (\ref{vertexbarphi}) for which the
mass-shell condition is
\begin{equation}
\alpha' k_1^2 = \alpha' k_2^2 =\ft12 -\ft12 \sum_i  \left[(\theta_{j (i)}^F)^2+\theta_j (1-\theta_j) \right]
= \frac 12 \sum_j \epsilon^F_{j (i)} ~\theta_j = - \alpha' M_i^2
\end{equation}
where
\begin{equation}
\label{eps}
 \epsilon_{j(i)}^F = \left\{  \begin{array}{ll}
 ~~1   &   i=j ~~,\\
 - 1   & i \not = j~~.
\end{array}
\right.
\end{equation}
For the closed string modulus, we use the vertices (\ref{vLR})
with the identifications (\ref{openclosed}). Thus, the amplitude (\ref{4ptbasic})
can be written as
\begin{equation}
\label{amplit}
\begin{aligned}
A_{j(i)}=& -\,\frac{\overline\phi^i(k_1)\phi^i(k_2)}{4\pi {\alpha'}} \int
\frac{dx_1 \,dx_2 \,d^2z}{dV_{\rm{CKG}}} \,{\rm e}^{-\ii\pi\alpha' k_L\cdot k_R}~ {\cal W} ~
(x_1-x_2)^{-1} \, (z-\bar{z})^{-2}
\\ & ~~\times
\left[
A_{\rm bos}^j(x_1,z,\bar z,x_2|\sigma_{\theta_j})
-
\, 2\alpha'k_L\cdot k_R
\, A_{\rm ferm}^j(x_1,z,\bar z,x_2|{\cal S}_{\theta^F_{j(i)}}) \right]
\end{aligned}
\end{equation}
where $A_{\rm bos}^j$ and $A_{\rm ferm}^j$ are the correlators
given in Eqs.~(\ref{boscorr}) and  (\ref{fermcorr4}) respectively,
and ${\cal W}$ is defined by
\begin{eqnarray}
{\cal W}&\equiv&
\big\langle \ex{\ii k_1\cdot X(x_1)}\,\ex{\ii k_L \cdot X(z)} \,\ex{\ii k_R\cdot
 X(\bar{z})}\,
 \ex{\ii k_2 \cdot X(x_2)} \big\rangle
 \,\prod_{j=1}^3\big[
\langle \sigma_{-\theta_j}(x_1)\sigma_{\theta_j}(x_2)
\rangle\,\langle{\cal S}_{-\theta^F_{j(i)}}(x_1){\cal S}_{\theta^F_{j(i)}}(x_2)
\rangle\big]\nonumber
 \\
 &=&(x_1-x_2)^{-1}\,\omega^{\alpha' (t+M_i^2)}\,(1-\omega)^{\alpha' s}
\label{calW}
\end{eqnarray}
in terms of the anharmonic ratio
\begin{equation}
\omega=\frac{(x_1-z)(\bar z-x_2)}{(x_1-\bar z)(z-x_2)}
~~~~~ (~|\omega|=1~)~~,
\label{wdef}
\end{equation}
and the Mandelstam variables
\begin{equation}
\label{kin1}
\begin{aligned}
s &= (k_1+k_2)^2 =(k_L+k_R)^2~~, \\
u &= (k_1+k_L)^2 =(k_2+k_R)^2~~,\\
t &= (k_1+k_R)^2=(k_2+k_L)^2 ~~,\\
\end{aligned}
\end{equation}
with
\begin{equation}
\label{kin2}
k_L^2 = k_R^2 = 0 ~~,~~ k_1^2 = k_2^2 = - M_i^2~~,~~ s+t+u = -2
M_i^2~~.
\end{equation}
In the following we keep the open strings on-shell, but we take the closed string off-shell.
If also the closed string were on-shell, we would have $u=t$ and $s=0$.
In our off-shell extension, instead, we retain the relation $u=t$ but
keep $s$ non-vanishing, {\it i.e.} we take $s = -2 (t + M_i^2)$.

Finally, in (\ref{amplit}) the open string punctures $x_1$ and
$x_2$ are integrated on the real axis, while the closed string
variable $z$ is integrated on the upper-half complex plane, modulo
the $\mathrm{Sl}(2,\mathbb{R})$ projective invariance which is fixed by the
Conformal Killing Group volume $dV_{\rm{CKG}}$. Using this fact,
one can show that
\begin{equation}
\frac{dx_1 \,dx_2 \,d^2z}{dV_{{\rm CKG}}}\,(x_1-x_2)^{-2}\,(z-\overline{z})^{-2}=
(1-\omega)^{-2} d\omega
\label{measure}
\end{equation}
and thus the amplitude $A_{j(i)}$ in (\ref{amplit}) becomes
\begin{equation}
\label{ampfin1b}
A_{j(i)}= - \frac{\overline\phi^i(k_1)\phi^i(k_2)}{4\pi\alpha'}\,{\rm e}^{-\ii\,\pi\alpha' s/2}\!
\int_{\mathcal{C}} d\omega~ \omega^{-\theta_j-\alpha' s/2}(1-\omega)^{\alpha' s-2}
\Big[1 - \theta_j (1-\omega) - \alpha' s \, \omega^{\theta_j-\theta^F_{j(i)}} \Big]~~.
\end{equation}
Notice that the original integral over $z$ takes into account all possible orderings of the
closed string insertion along the open string boundary. In the $\omega$-variable
this translates into a closed integral $\mathcal{C}$
along the unit circle $|\omega|=1$ clockwise oriented\footnote{To see that
the unit circle $\mathcal{C}$ is clockwise oriented we can consider the definition
of $\omega$ in Eq.~(\ref{wdef})
and take the limits $x_1\to \infty$ and $x_2\to 0$, so that $\omega\to \overline z/z$.
Since $z\in \mathbb{H}_+$, we easily see that $\omega\to {\rm e}^{-2\ii\varphi}$
with $0\leq\varphi<\pi$, and thus $\mathcal{C}$ is covered
clockwise.}. The integrand in (\ref{ampfin1b}) has a branch cut along the positive real axis and
thus the contour $\mathcal{C}$ must be deformed in order to circumvent the cut
singularity.
So we have to perform the integration just below the cut for $\omega\in [0,1]$ and then subtract
the contribution from above the cut for $\omega\in \ex{-2 \pi \ii} [0,1]$.
Using the definition of the Euler $B$-function,
the amplitude (\ref{ampfin1b}) takes the form
\begin{equation}
\begin{aligned}
A_{j(i)} &= \,-\,
\frac{\ii\,\overline\phi^i(k_1)\phi^i(k_2)}{2\pi \alpha' }\,{\rm e}^{\ii\pi\theta_j}\,
\sin\Big[\pi\big(\theta_j+{\alpha's}/{2}\big)\Big]\,
\Bigg\{B(1-\theta_j-\alpha's/2,\alpha's-1) \\
& ~~~~~~- \theta_j\,
B(1-\theta_{j}-\alpha's/2,\alpha's)-\alpha's\,B(1-\theta^F_{j(i)}-\alpha's/2,\alpha's-1)
\Bigg\}\\
&=\,-
\frac{\ii\,\overline\phi^i(k_1)\phi^i(k_2)}{4\pi \alpha'}\,\epsilon_{j(i)}^F
\,{\rm e}^{\ii\pi\theta_j}
\,\sin\Big[\pi\big(\theta_j+{\alpha's}/{2}\big)\Big]\,\,
\frac{\Gamma(\alpha' s+1) \Gamma(1 - \theta_j - \alpha'{s}/{2})}{\Gamma(1 - \theta_j
+ \alpha'{s}/{2})}
\label{ampgen1}
\end{aligned}
\end{equation}
where $\epsilon_{j(i)}^F$ are the signs introduced in (\ref{eps}).

In order to extract information about the low energy effective action, a few further steps
must be performed. First we have to expand our result (\ref{ampgen1}) in powers of
$\alpha' s$, then take the limit $\alpha' \rightarrow 0$ keeping the mass $M_i$ fixed: in this
way the first two terms in such expansion yield the mass and the kinetic terms
for the scalar fields, if we use the relation
$s=2k_1\cdot k_2-2M_i^2$.
Then, we get
\begin{equation}
\label{amp1shell}
A_{j(i)} = -\frac{\ii\,\overline\phi^i(k_1)\phi^i(k_2) }{4\pi\alpha'}\,
\epsilon^F_{j(i)}\,\ex{\ii\pi\theta_j} \sin (\pi \theta_j) (1-\frac{1}{2}\,\alpha' s\,
\rho_{j}) + \,{\cal O}\left(\alpha' s^2\right)
\end{equation}
with
\begin{equation}
\label{rho1}
\rho_j =\psi(1-\theta_j)+ \psi(\theta_j) + 2 \gamma_E ~~.
\end{equation}
Here $\gamma_E$ is the Euler-Mascheroni constant and
$\psi(x)= d \,{\rm ln} \Gamma(x)/dx$. In writing $\rho_j$ as in
(\ref{rho1}),
we have used the identity $ \psi(1-\theta_j) =
\psi(\theta_j) + \pi \cos(\pi\theta_j)/\sin(\pi\theta_j)$.
Notice that the result (\ref{amp1shell}) holds independently
from supersymmetry, {\it i.e.} it is valid
both for $M_i=0$ and $M_i \not = 0$, and is {\it exact} in $\alpha'$.

Proceeding as above, one can show that the amplitude ${A}_{{\overline j} (i)}$
is given by the same expression (\ref{amp1shell}) with
$\ex{\ii\pi\theta_j}$ replaced by $\ex{-\ii\pi\theta_j}$.

\vskip 0.5cm
\subsection{Identifying the K\"ahler metric}

In this section we finally extract the K\"ahler metric $K_{ii}$ from the string
amplitude  ${\cal A}_{m\bar \phi^i \phi^i}$. In order to achieve this goal,
we show that the string amplitude can be written in the form (\ref{lagder}) from
which the expression of the K\"ahler metric can be read off.
Let us define for convenience
\begin{equation}
h_{j(i)}=\frac{\overline\phi^i(k_1)\phi^i(k_2) }{4\pi\alpha'
}\,\epsilon^F_{j(i)}\,  \big(1-\frac{1}{2} \alpha'\, s\,
\rho_{j}\big)
\label{hji}
\end{equation}
and introduce the matrix
\begin{equation}
{\cal H}_{(i)}= \left(
\begin{array}{cc}
h_{j(i)}\,\delta^j_{~k} & 0 \\
  0 & -h_{\bar j(i)} \,\delta^{\bar j}_{~\bar k}\\
\end{array}
\right)~~~~{\rm with}~~~~h_{\bar j(i)}={\overline h}_{j(i)}~~.
\label{calH}
\end{equation}
Then, after simple manipulations one sees that Eq.~(\ref{calAij}) can be rewritten as
\begin{equation}
{\cal A}_{(i)}=
\frac12 \,{\cal G}^{-1}\, ({\cal R}^{-1}-1)\, {\cal H}_{(i)} ~~.
\end{equation}
Plugging this back into Eq.~(\ref{deriv0}), one finally finds
\begin{equation}
\begin{aligned}
{\cal A}_{m\bar \phi^i \phi^i} &={\rm tr} \left[
{}^t {\cal E}^{-1}\, \frac{\partial}{\partial m} (G-B)~R_0~ {\cal E}^{-1} ~ {}^t{\cal
A}_{(i)}
\right]\\
&= \frac12 \, {\rm tr} \left[
{}^t {\cal E}^{-1}\, \frac{\partial}{\partial m} (G-B) ~(R_\pi -R_0)~
{\cal E}^{-1}{\cal H}_{(i)}~{\cal G}^{-1} ~
\right]\\
&= \frac12\sum_{j=1}^3\Bigg[\left(\ECB\,
 \GB^{-1} \,\frac{\partial(\GB - \BB)}{\partial m}\, \left(\RB_\pi -
 \RB_0\right)
\,\ECB^{-1}\right)_{jj}-{\rm h.c.}\Bigg]\,
h_{j(i)}~~.
\label{ampfinjac}
\end{aligned}
\end{equation}
At this point the crucial observation is that the term multiplying $h_{j(i)}$ in the above
expression can be written as a total derivative with respect to $m$.
This fact follows from the non-trivial identity (\ref{dmt8}),
whose proof is presented in Appendix \ref{app:jac}.
Using this identity in (\ref{ampfinjac}), we find
\begin{equation}
\label{ah}
{\cal A}_{m\bar{\phi}^i \phi^i} = 2 \pi{\ii}\, \sum_{j=1}^3 h_{j(i)}\,
\frac{\partial \theta_j}{\partial m}~~.
\end{equation}
Then, inserting the relation $s=2k_1\cdot k_2- 2M_i^2$
in the explicit expression of $h_{j(i)}$ given in (\ref{hji}), we get
\begin{equation}
\begin{aligned}
2\pi h_{j(i)} &= \epsilon^F_{j(i)}\, \left[ \frac{1}{2 \alpha'}
-\frac{1}{2} \, \big(k_1\cdot k_2-M_i^2\big) \,\rho_j \right]\,\overline\phi^i(k_1)\phi^i(k_2) \\
&= \Bigg[\frac{\epsilon^F_{j(i)}}{2 \alpha'}
+ (k_1 k_2-M_i^2)
 \frac{d}{d\theta_j} \ln
 \Big( {\rm e}^{-2\gamma_E\theta_j}\,\frac{\Gamma(1-\theta_j)}{\Gamma(\theta_j)} \Big)^{
\epsilon^F_{j(i)}/2}\Bigg]\,\overline\phi^i(k_1)\phi^i(k_2)
\end{aligned}
\end{equation}
where in the last step we have used the definition (\ref{rho1}) for
$\rho_j$. From the analysis of the spectrum we know the value of the
tree-level mass $M_i^2 = -\frac{1}{2\,\alpha '}\,\sum_j
\epsilon^F_{j(i)}\, \theta_j$, which allows to interpret the first term in
the equation above as $-\partial M_i^2/\partial\theta_j$. Notice that this
is a way to fix unambiguously the overall normalization of the string
amplitude and thus also the power of the K\"ahler metric below. At this
point we can use Eq.~(\ref{ah}) to write the amplitude ${\cal
A}_{m\bar{\phi}^i \phi^i}$ in the form of~\Eq{lagder} and read the
explicit form of $K$
\begin{equation}
K_{ii}(\theta) = {\rm e}^{2 \gamma_E\,\alpha' M_i^2}\,
\prod_{j=1}^3
\left(\frac{\Gamma(1-\theta_j)}{\Gamma(\theta_j)}\right)^{\epsilon^F_{j (i)}/
2}~~.
\label{gkahlers}
\end{equation}
Formula (\ref{gkahlers}) is the main result of this paper and, as we
discuss below, it generalizes previous results in the literature.  It
displays the full moduli dependence of the K\"ahler metric of the
chiral matter coming from the $\theta$-twisted open strings and holds
for an arbitrary brane setup in presence of generic {\it
  non-commuting} fluxes. Notice that the result (\ref{gkahlers}) holds
independently on whether supersymmetry is preserved or broken, {\it
  i.e.} it is valid even when $M_i\neq 0$.  Remarkably, the K\"ahler
metric is always determined by a simple function of the twists
$\theta_i$.  It is worth stressing that in this derivation it is
useful to keep $M_i\neq 0$ in order to fix the overall normalization
of the string amplitude, including the sign, and hence to determine in
the end the exact power in (\ref{gkahlers}).

Notice that Eq.~(\ref{gkahlers}) is {\it exact} in $\alpha'$. The
field-theory result is obtained by taking
the limit $\alpha' \rightarrow 0$ with $M_i^2$ fixed. In
this limit the exponential vanishes and the K\"ahler metric entering
the field-theory Lagrangian finally reads
\begin{equation}
K^{(0)}_{ii}=\lim_{\alpha'\to 0} K_{ii}(\theta) =
\sqrt{\frac{\Gamma(1- \theta^{(0)}_i)}
{\Gamma(\theta^{(0)}_i)}}\,\prod_{j\not= i}  \sqrt{\frac{\Gamma(\theta^{(0)}_j)}
{\Gamma(1 - \theta^{(0)}_j)}}
\label{gkahler}
\end{equation}
where $\theta^{(0)}_j=\lim_{\a'\to 0} \theta_j$ as in
(\ref{thetaalpha}), and the signs $\epsilon^F_{j(i)}$ for the scalar $\phi^i$
have been made explicit.

Some comments are in order at this point. First we notice that if we
start from a non-supersymmetric set of $\theta_i$'s, the only way to
decouple one of the twisted scalars from the string scale is to
suppose that the $\theta_i^{(0)}$'s satisfy a supersymmetric
constraint, as is clear from the mass formula (\ref{mass2}).  This
means that one is considering a particular point in
Fig.~\ref{tetrahedron} which is at a ``stringy'' distance from a given
supersymmetric configuration, in such a way that one scalar can
survive in the field theory limit $\alpha'\to 0$ with a finite mass.
As we will discuss in the next section this breaking
can be interpreted at the field theory level as coming from a non-vanishing
v.e.v. of a Fayet-Iliopoulos term.

Notice that in our derivation
we have chosen a set of conventions for which the ${\cal N}=4$ supersymmetric point
included in our $\theta$-space is the vertex ${\cal O}$ of the
tetrahedron in Fig.~\ref{tetrahedron}.
However, our results hold for any other choice. For instance, we
could repeat the above analysis with different
conventions and consider a field theory
where the starting point is another ${\cal N}=4$ vertex of the tetrahedron in
Fig.~\ref{tetrahedron}. In this case, the scalar becoming massless on the outer
wall $({\cal ABC})$ would now enter the low energy effective spectrum and
its K\"ahler metric would be
\begin{equation}
\label{finmetsym}
K^{(0)}= \prod_{j=1}^3  \left(\frac{\Gamma(1-\theta^{(0)}_j)}{\Gamma(\theta^{(0)}_j)}
\right)^{1/2}~~.
\end{equation}

This is the scalar that is usually considered in the literature. However,
we point out that the exponent in Eq.~\eq{finmetsym} has a different sign as
compared to previous findings, but it agrees with the result of
Ref.~\cite{Font:2004cx}. In the Heterotic computations~\cite{Dixon:1989fj} the K\"ahler metric for
the scalar fields is derived from a four point amplitude on the sphere.
One may wonder why in models with open strings it is possible to derive
this result from a three point function and, conversely, what r\^ole a four
point function would have in this context. However, at the CFT level, the
insertion of a closed string on a disk is equivalent to the insertion of
two open vertices. Thus the Koba-Nielsen integrals are those also
considered in Ref.~\cite{Dixon:1989fj}. Of course, the space-time interpretation
and kinematics are those of a three point function. Thus, to get a
meaningful result, it is important to give a prescription to continue the
string amplitude off-shell, as we do, at least in the field theory limit. It would
be very interesting to check this off-shell prescription by computing disk
diagrams with the insertion of two moduli vertices and see whether the
results are consistent with~\Eq{gkahlers}. This is a challenging
computation and some preliminary results were presented
in~\cite{Lust:2004cx} for the factorized and commuting case. The authors
of~\cite{Lust:2004cx} suggested that the differential equations derived
from the three point function should actually be modified to agree with
the results coming from higher point amplitudes. Here we seem to have no
room for modifications of this type. We present a check of this in
Appendix~\ref{app:unt}; there we focus on the untwisted scalars where it
is possible to compare the result with the Born-Infeld action and we find
complete agreement. So we believe that our off-shell prescription is able
to capture the full NS-NS moduli dependence of the metric for all scalar
fields.

\vskip 0.5cm
\subsection{Commuting cases}

To make contact with previous results in the literature, here we illustrate
our results in the simplest situation where the flux and the
reflection matrices $R_\sigma$ commute.
Representatives of this commuting case are the branes with ``diagonal
fluxes '' discussed in Section \ref{subsecn:fac_diag}.
Using the results derived there (but dropping for simplicity the
index $i$ labeling the three torii $\mathcal{T}^2_i$), one can explicitly verify that derivatives
of the twist parameter satisfy Eq.~(\ref{dmt8}). Indeed, from
(\ref{theta2d}) it follows that
\begin{equation}
\label{dertdir2}
 2\pi\ii\,\frac{\partial\theta}{\partial T}  =
\frac{f_\pi - f_0}{(T - f_\pi)(T - f_0)}~~.
\end{equation}
Using Eqs.~(\ref{GB2}) -- (\ref{E2d}) it is not difficult
to check that the general relation (\ref{dmt8}) correctly reproduces
Eq. (\ref{dertdir2})
(see Appendix \ref{subapp:jac2} for further details).

As is well-known, a T-duality along the $y$ direction of the torus
$\mathcal{T}^2$ corresponds to the exchange $T\leftrightarrow
-1/U$, and the magnetized branes of the type IIB theory become branes of type IIA
intersecting at angles.
A careful analysis of the boundary conditions
(\ref{bc1}) reveals that under this T-duality the reflection matrices
(\ref{Rs2d}) transform into
\begin{equation}
\RCB'_\sigma =
\begin{pmatrix}
0 & - \frac{1+ \overline U\, f_\sigma}{1+ U\, f_\sigma}
\\
- \frac{1+ U\, f_\sigma}{1+ \overline U\, f_\sigma} &0
\end{pmatrix}
~~.
\label{Rdual}
\end{equation}
Clearly $\RCB'_0 $ and $\RCB'_\pi$ commute with each other.
These T-dual reflection matrices depend
on the complex structure $U$ and the quantized magnetic fluxes
$f_\sigma$, but are independent of the K\"ahler modulus $T$, in contrast to the
original matrices $\RCB_\sigma$ of Eq. (\ref{Rs2d}).
The monodromy matrix $\RCB' = \big({\RCB'}_\pi\big)^{-1} \RCB'_0$ of the T-dual theory
is of the form
$\RCB' = \diag \left(\ex{2\pi\ii\theta'},\ex{-2\pi\ii\theta'}\right)$ with
\begin{equation}
\label{theta2dT}
\ex{2\pi\ii\theta'} = \frac{1 + U \,f_\pi}{1 + \overline U \,f_\pi}
\, \frac{1 + \overline U \, f_0}{1 + U \, f_0}~~,
\end{equation}
which is the direct T-dual transform of \Eq{theta2d}.
In this case the twist $\theta'$ represents the intersecting angle
between the two D-branes to which the open string is attached. If we take one of the
branes to lie on the $x$ axis, i.e. if we set $f_0 = 0$, then \Eq{theta2dT} can be simply
rewritten as
\begin{equation}
\tan(\pi\theta') = \frac{U_2\,p}{
q +U_1 p}
\label{thetai4}
\end{equation}
where the quantization condition $f_\pi=p/q$
has been used. This is the usual relation of the angle between two D-branes
with the complex structure moduli of the two-dimensional torus
$\mathcal{T}_2$ in which they intersect,
a relation that can be easily understood and derived also in geometrical terms.
From Eq. (\ref{thetai4}) it follows that
\begin{equation}
2\pi\ii\,\frac{\partial\theta'}{\partial U} =
\frac{2 p}{(q+U\,p)}~~,
\label{jac}
\end{equation}
which again agrees with the general result (\ref{dmt8}).
Notice that indeed Eqs. (\ref{dertdir2}) and (\ref{jac}) are related by the T-duality
map $T \Leftrightarrow -\frac{1}{U} $. More generally, under a T-duality transformation
$\bar{X}^M=T^M{}_N \bar{X}^N$, it can be shown that the flux matrices
transform as \cite{Ooguri:1996ck}
\begin{equation}
\left[R_\sigma'(m')\right]^M{}_N =\left[R_\sigma(m(m'))\right]^M{}_P \, T^P{}_N
\end{equation}
where $m$ and $m'$ the T-dual moduli and the $d\times d$ matrix $T^M{}_N$
satisfies $T_{MN}=T_{NM}$ and $T^2=1$. The dependence on this matrix $T$ cancels out in the
monodromy matrix $R$ and therefore open
string twists in T-dual theories are simply related by replacing
$m \Leftrightarrow m'$. The case of the T-duality in the $y$ direction of the two torus
$\mathcal{T}^2$ discussed above is just an explicit example of this
more general statement.

\vskip 0.5cm
\subsection{Supersymmetry breaking by $D$- and $F$-terms}

In presence of generic fluxes (or angles) supersymmetry may be
broken by $D$- and $F$-terms. Here we would like to analyze
explicitly from a string theory point of view these mechanisms
starting from the one produced by $D$-terms.

Let us then compute the v.e.v. of the auxiliary fields $D$
of the gauge vector multiplet for our system of magnetized
branes with generic fluxes. Since the chiral matter arises from open strings
stretched between two (stacks of) D9-branes, we should consider
both the $D$ field of the gauge multiplet for the branes at $\sigma=0$ and the
$D$ field on the branes at $\sigma=\pi$, and then focus on their
respective $\mathrm{U}(1)$ parts which are the only ones that can
get a v.e.v. In particular we should compute from string diagrams the difference
\begin{equation}
\langle D \rangle_\pi-\langle D \rangle_0
\label{dterm}
\end{equation}
and show that, as expected, it corresponds to a mass for the twisted chiral matter.
Just like we did for the K\"ahler metric, we will actually compute the
derivative of the above quantity with respect to a closed string
modulus $m$, rather than the v.e.v.'s themselves. More precisely
we consider a disk amplitude between a vertex operator $V_D$ for the
auxiliary field $D$ and a closed string vertex operator $W_m$ for the
modulus $m$, and read from it the v.e.v. of the $D$ fields according to
\begin{equation}
\label{vevD}
\mathcal{A}_{mD} \equiv
\langle W_m  \,V_D\rangle_\pi
-\langle W_m  \,V_D\rangle_0 =\ii\,\frac{\partial}{\partial m}\Big(
\langle D \rangle_\pi-\langle D \rangle_0
\Big)
\end{equation}
where the subscripts $0$ and $\pi$ on the string correlators
indicate that the appropriate boundary conditions for the branes at $\sigma=0$ and
$\sigma=\pi$ should be enforced.
Auxiliary fields are realized in
string theory in terms of non-BRST invariant operators in the 0-superghost picture
(see, for example, Ref.~\cite{Dine:1987gj} for details and
Ref.~\cite{Billo:2005jw} for some recent applications in mixed open/closed string amplitudes) given by
\begin{equation}
V_D(z) = \frac12 \,\xi_{(i)MN}\,:\Psi^M(z)  \Psi^N(z):
\label{vertexD}
\end{equation}
where $\xi_{(i)}$ is the imaginary part of the K\"ahler form of the
internal torus. The label $(i)$ specifies along which ${\cal N}=1$ supersymmetry, out
of the starting ${\cal N}=4$, the auxiliary field under consideration is aligned\footnote{Here
we use the same notation already adopted to distinguish the three faces of the tetrahedron
of Fig.~\ref{tetrahedron}, which correspond to three different $\mathcal{N}=1$ supersymmetries.
The fourth supersymmetry associated to the outer face of the tetrahedron is out
of the present discussion, but, as we have already seen,
it could be incorporated without any problem by simply changing our conventions.}.
In the complex basis (\ref{diagonalR1}) we have
\begin{equation}
{}^t {\cal E}^{-1}\,  \xi_{(i)} \, {\cal E}^{-1}=\left(
\begin{array}{cc}
  0 & {\epsilon}_{j(i)}^F\, \delta_{j\bar k} \\
  -{\epsilon}_{j(i)}^F \,\delta_{\bar j k} & 0 \\
\end{array}
\right)
\label{polD}
\end{equation}
where are the signs introduced in Eq.~(\ref{eps}).
In writing $\xi_{(i)}$ in this form we use the fact that in the
orthonormal basis the metric of the torus is of the form (\ref{Gcal}) and
rearrange rows and columns in order to ensure that the twists $\theta_i$'s
are all positive.

Since the vertex $V_D$ is in the 0-superghost picture, we need to
take the closed string vertex $W_m$ in the ($-1,-1$) picture,
where it is given by Eq. (\ref{vm}) with
\begin{equation}
V_L^M(z)= \frac{1}{\sqrt 2}\,
{\rm e}^{-\varphi_L(z)}\, \Psi_L^M(z)~~~~{\rm and}~~~~
V_R^M(\overline z)= \frac{1}{\sqrt 2}\,
{\rm e}^{-\varphi_R(\overline z)}\, \Psi_R^M(\overline z)~~.
\label{vm-1}
\end{equation}
As already mentioned, the amplitude $\mathcal{A}_{mD}$ receives contributions
from insertions in the disks at $\sigma=0$ and $\sigma=\pi$ with boundary
conditions parameterized by the reflection matrix $R_\sigma$,
so that the identifications of the left and right moving parts of the closed
string with the propagating (untwisted) open string are
\begin{equation}
V_L^M(z) = V^M(z)~~~~{\rm and}~~~~V_R^M(\overline z) = (R_\sigma)^M_{~N}V^N(\overline z)
~~.
\end{equation}
Collecting all pieces one finds
\begin{equation}
\begin{aligned}
{\cal A}_{mD} &=
\frac{1}{16\pi \alpha'} \,\xi_{(i)PQ}
\left[\frac{\partial}{\partial m} (G-B)\,(R_\pi-R_0)\right]_{MN}
\\
&~~~~\times\int\frac{dx \,d^2z}{dV_{\rm{CKG}}}
~\Big\langle {\rm e}^{-\varphi(z)}\,\Psi^M(z)~ {\rm e}^{-\varphi(\overline z)}\,
\Psi^N (\overline{z})~:\Psi^P \Psi^Q:(x) \Big\rangle \\
&= \frac{1}{8\pi \alpha'} \, {\rm tr}\,\left[ G^{-1}\, \xi_{(i)} \, G^{-1}
\frac{\partial}{\partial m} (G-B)\,(R_\pi-R_0)\right]~~.
\label{resad}
\end{aligned}
\end{equation}
Notice that the full dependence on world-sheet positions cancels in the integrand
in agreement with the $\mathrm{Sl}(2,\mathbb{\mathbb{R}})$ invariance.
That only the $U(1)$ part of $V_D$ contributes to ${\cal A}_{mD}$ is clear since
this amplitude is proportional to the trace of the Chan-Paton factor carried by
the $D$-vertex.
Finally, using (\ref{polD}) to rewrite the polarization $\xi_{(i)}$ in the orthonormal basis
and exploiting the non-trivial identity (\ref{dmt8}), one finds
\begin{equation}
\begin{aligned}
{\cal A}_{mD}
&=-\frac{1}{8\pi \alpha'}\,\sum_{j=1}^3
\Bigg[\left(\ECB\,
 \GB^{-1} \,\frac{\partial(\GB - \BB)}{\partial m}\, \left(\RB_\pi -
 \RB_0\right)
\,\ECB^{-1}\right)_{jj}-{\rm h.c.}\Bigg]\,{\epsilon}_{j(i)}^F
\\
&= -\,\frac{\ii}{2 \alpha'}\,\sum_{j=1}^3 {\epsilon}_{j(i)}^F
\,\frac{\partial\theta_j}{\partial m} \,=\,
\ii\,\frac{\partial M_i^2}{\partial m}~~.
\end{aligned}
\end{equation}
Comparing with Eq. (\ref{vevD}), we see that indeed
$M_i^2=\langle D \rangle_\pi-\langle D \rangle_0$, thus proving
that the twisted scalars $\phi^i$ become massive when the
$D$ fields acquire a v.e.v. This calculation shows also in a very explicit way
that the subleading terms $\epsilon_i$ in the open string twists, defined
in (\ref{thetaalpha}), which responsible for the scalar mass, have the
interpretation of Fayet-Iliopoulos parameters in the effective low-energy theory.

In a similar way one can compute also the $F$-terms, {\it i.e.} the v.e.v. of the auxiliary
fields $F^i$ and $F^{\bar i}$ of the adjoint chiral multiplets of the untwisted sector.
Their corresponding vertex operators are of the form
\begin{equation}
V_{F^i}(z) = \frac12 \,\zeta^{i}_{MN}\,:\Psi^M(z)  \Psi^N(z):
~~~~{\rm and}~~~~
V_{F^{\bar i}}(z) = \frac12 \,\zeta^{\bar i}_{MN}\,:\Psi^M(z)  \Psi^N(z):
\label{vertexF}
\end{equation}
where the polarizations $\zeta^i$ and $\zeta^{\bar i}$, in the
complex orthonormal basis, are
\begin{equation}
{}^t {\cal E}^{-1}\,  \zeta^{i} \, {\cal E}^{-1}\sim\left(
\begin{array}{cc}
  0 & 0 \\
 0& \delta^{i\bar j}\,\epsilon_{\bar j\bar k\bar \ell} \\
\end{array}
\right)
~~~~{\rm and}~~~~
{}^t {\cal E}^{-1}\,  \zeta^{\bar i} \, {\cal E}^{-1}\sim\left(
\begin{array}{cc}
 \delta^{\bar i j}\,\epsilon_{ jk\ell} & 0 \\
 0&  0 \\
\end{array}
\right)~~.
\label{polF}
\end{equation}
Notice that unlike the polarization (\ref{polD}) of the $D$ vertex
operator, these polarizations have non vanishing entries
in the diagonal blocks when they are expressed in the orthonormal
frame.

The v.e.v. of $F^i$ and $F^{\bar i}$ can be obtained from the
string amplitudes $\mathcal{A}_{mF^i}$ and $\mathcal{A}_{mF^{\bar
i}}$, which have the same form as (\ref{resad}) but with  $\xi_{(i)}$ replaced
by $\zeta^i$ and $\zeta^{\bar i}$. Due to the structure of these
polarizations, we immediately see that in the interesting case where fluxes
are of type $(1,1)$ and $\partial_m (G-B)$ is block off-diagonal,
there is no F-term since the trace in (\ref{resad}) vanishes. In
this way we see that fluxes of type (1,1) do not give rise to any
$F$-term and hence do not break supersymmetry. On the contrary,
fluxes of the type (2,0), which correspond to a $\partial_m (G-B)$
with non-vanishing entries also in the diagonal blocks, do produce
an non-vanishing $F$-term amplitude and hence induce a
non-vanishing v.e.v. for these auxiliary fields.
{F}rom the structure of the vertex operators (\ref{vertexF}) it is
easy to realize that in the field theory limit $F^i$ and $F^{\bar i}$ do not couple
to the chiral fields of the twisted sector, and thus the presence
of a non-vanishing $F$-term does not break supersymmetry there.
However, supersymmetry will be broken by these $F$-terms in other sectors, for
example in the bulk.

\vskip 0.8cm
\sect{Relation with the Yukawa Couplings}
\label{sec:yukawa}

\begin{figure}[htp]
\hspace{.5cm}
\begin{center}
\begin{picture}(0,0)%
\includegraphics{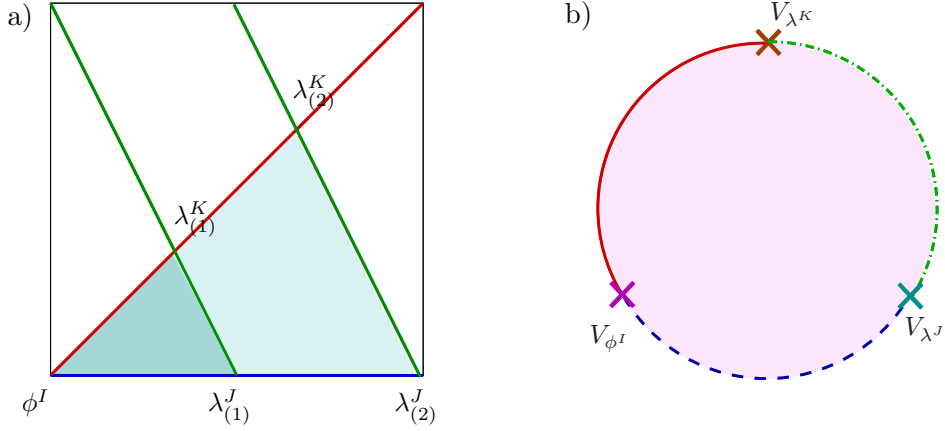}%
\end{picture}%
\input yuk_1.pstex_t
\end{center}
\caption{\small a) Classical contributions $\mathcal{W}_{IJK}$ to the
Yukawa couplings in intersecting D-brane models.  b) Quantum
contributions $\mathcal{A}_{IJK}$ are given by string correlators}
\label{fig:yuk}
\end{figure}

The Yukawa couplings among the fields arising from intersecting or
magnetized brane worlds admit a nice stringy
description~\cite{Cremades:2003qj}, which represents actually one of
the strong points of such constructions. We focus on the couplings
between chiral fermions and scalars all arising from twisted strings.
In this stringy description, the couplings appearing in the Yukawa
terms of the effective action have the form
\begin{equation}
\label{yuk1}
Y_{IJK} = \mathcal{A}_{IJK}\,\mathcal{W}_{IJK}~~,
\end{equation}
where $I,J,K$ are generic indices denoting the various scalars and
fermions, which we will specify better in the cases we are actually
concerned with. Here $\mathcal{W}_{IJK}$ we denote \emph{classical}
contributions, which in the case of intersecting
branes~\cite{Cremades:2003qj,Cvetic:2003ch,Abel:2003vv,Lust:2004cx}
are given by world-sheet instantons bordered by the intersecting
branes\footnote{The world-sheet instanton contributions have obviously
  a counterpart in the magnetized brane models, which is discussed for
  instance in \cite{Cremades:2004wa}.}.  In this context, since the
replica families of fields arise from multiple intersections of the
branes, the different areas of the minimal world-sheet connecting
different intersections provide naturally an exponential
hierarchy of couplings, see Fig.~\ref{fig:yuk}a).

The (string) quantum contributions $\mathcal{A}_{IJK}$ to the couplings
are instead provided by the correlator of the twisted emission
vertices of a scalar $V_{\phi^I}$ (from the NS sector of a twisted
string) and of two fermions $V_{\lambda^J}$ and $V_{\lambda^K}$, from
the R sector of two other twisted strings, see Fig.~\ref{fig:yuk}b)
\begin{equation}
\label{yuk5}
\mathcal{A}_{IJK} =
\langle V_{\phi^I} V_{\lambda^J} V_{\lambda^K}\rangle
\propto \int \frac{dx_1 dx_2 dx_3}{dV_{\mathrm{CKG}}}
\langle V_{\phi^I}(x_1)  V_{\lambda^J}(x_2) V_{\lambda^K}(x_3)\rangle~~.
\end{equation}
A three-point CFT correlator is determined from conformal invariance up
to a constant; since the vertices have conformal dimension 1, this
structure constant coincides directly with the string amplitude
\begin{equation}
\label{yuk5.1}
\langle V_{\phi^I}(x_1)  V_{\lambda^J}(x_2) V_{\lambda^K}(x_3)\rangle
= \frac{\mathcal{A}_{IJK}}{\prod_{a,b=1}^3 (x_a - x_b)}~~.
\end{equation}
The world-sheet dependence from the $x_a$, indeed, just cancels in the
amplitude \Eq{yuk5} against the Jacobian to gauge-fix
$\mathrm{SL}(2,\mathbb{R})$ invariance.

We consider $\mathcal{N}=1$ configurations, in which
supersymmetry may be broken, as we have just seen, by the presence of
D-terms. In $\mathcal{N}=1$ theories, the Yukawa couplings are encoded
in the superpotential. In truth, our effective action is an
$\mathcal{N}=1$ supergravity, and beside the twisted matter multiplets
$\Phi^I$ we have matter multiplets originating from the closed string
sector, including the moduli scalars $m$.  As we did for the K\"ahler
potential, though, we presently consider the moduli $m$ as fixed and
expand the superpotential in the open string multiplets $\Phi^I$. The
cubic level of this expansion
\begin{equation}
\label{yuk1.1}
W = W_{IJK}(m)\,\Phi^I\Phi^j\Phi^K~~,
\end{equation}
displays the holomorphic couplings $W_{IJK}$ when the moduli are
written in the appropriate complex basis. These couplings govern the
Yukawa terms involving one scalar and two fermions from these
multiplets.  They, however, do not directly represent the physical
Yukawa couplings $Y_{IJK}$ because in the $\mathcal{N}=1$ Lagrangian
the chiral multiplet fields have non-canonical kinetic terms involving
the K\"ahler metric $K_{IJ}(m)$.  To read off the physical
couplings\footnote{We assume here that the K\"ahler metric is diagonal
  in the space of the $\Phi^I$, which is indeed the case for the
  twisted matter we consider.}  we have to rescale the fields: $\Phi^I
\to (K_{II})^{-1/2}\, \Phi^I$, see the discussion before
\eq{rescaling}, getting
\begin{equation}
\label{yuk2}
Y_{IJK} = \Big[K_{II}K_{JJ}K_{KK}\Big]^{-1/2}\,W_{IJK}~~.
\end{equation}

For $\mathcal{N}=1$ effective theories realized in Heterotic string
compactifications, a powerful non-renormalization
theorem~\cite{Witten:1985bz} asserts that the superpotential $W$ gets
no perturbative $\alpha'$ corrections. It is likely that the same
non-renormalization property holds also in the brane-world context. If
this is the case, we should identify the holomorphic couplings
$W_{IJK}$ with the classical world-sheet instanton contributions
\begin{equation}
\label{yuk3}
W_{IJK} = \mathcal{W}_{IJK}~~,
\end{equation}
since these contributions depend non-perturbatively on $\a'$:
$\mathcal{W}_{IJK} \sim {\rm e}^{S/\a'}$. On the other hand, the
string amplitude $\mathcal{A}_{IJK}$ with three twisted vertices can
be certainly expanded perturbatively in $\a'$ and so can not
contribute to the form of the superpotential. It then follows from
Eqs.~\eq{yuk1} and~\eq{yuk2} that $\mathcal{A}_{IJK}$ should factorize
in term of the K\"ahler metrics for the involved fields, namely
\begin{equation}
\label{yuk4}
\mathcal{A}_{IJK} = \Big[K_{II}K_{JJ}K_{KK}\Big]^{-1/2}~~.
\end{equation}
This remarkable statement should be checked against the direct
computation of the string correlator $\mathcal{A}_{IJK}$. Let us
analyze this problem and, to begin with, let us recall which chiral
multiplets we consider and set up a convenient notation.

As we discussed in \secn{secn:spectrum}, an open string stretching
between two different D-branes is characterized by the eigenvalues
$\theta_i$ ($i=1,2,3$) of its monodromy $R(\theta)$.  It contains in
its NS spectrum three different scalars $\phi^i$ which can be retained
in the effective theory also for non-trivial values of the twists.
These are the states of \Eq{scalars2}; their mass is given in
Eqs.~(\ref{scalars2}-\ref{mass2}), and the corresponding emission
vertices in \Eq{vertexphi}. For $\sum_{j\not=i} \theta_j - \theta_i =
0$, the scalar $\phi^i$ is massless and sits in a chiral
$\mathcal{N}=1$ multiplet with the massless chiral fermion $\lambda$
from the R sector.  The latter is present for any value of the
$\theta$'s, and its emission vertex was written in \Eq{vertexR}.

In a given model of magnetized or intersecting D-branes, there are
various types of branes, and many open strings sectors associated to
various pairs of different D-branes. These sectors are distinguished
by the corresponding monodromy $R(\theta)$ and its eigenvalues, the
twists $\theta_j$. We can thus label\footnote{That is, the index $I$ is a
shortcut for $(\theta,i)$, in this case: the twist singles out which open string
sector and the index $i=1,2,3$ which scalar component \Eq{scalars2} we refer
to.}  the chiral multiplets as $\Phi^i_\theta$.  The
corresponding K\"ahler metric is evidently diagonal in the space of
different open string sectors, and also with respect to the type $i$
of scalars under consideration. We choose for it the notation
$K_{ii}(\theta)$. The string amplitude computing the quantum part of
the Yukawa couplings among three such chiral multiplets
$\Phi^i_\theta$, $\Phi^j_\nu$ and $\Phi^k_\omega$ is
\begin{equation}
\label{yuk5.2}
\mathcal{A}_{ijk}(\theta,\nu,\omega) =
\langle V_{\phi^i_\theta} V_{\lambda_\nu} V_{\lambda_\omega}\rangle
\end{equation}
and it is encoded in the conformal correlator of the vertices as
indicated in \Eq{yuk5.1}. We restrict ourselves to the coupling
between multiplet of the same type $i=j=k$, say for instance $i=1$. We
can then simplify further the notation, writing $K(\theta)$ for the
K\"ahler metric $K_{11}(\theta)$ and $\mathcal{A}(\theta,\nu,\omega)$
for the quantum Yukawa amplitude
$\mathcal{A}_{111}(\theta,\nu,\omega)$.

\vskip 0.5cm
\subsection{The factorized case}
\label{subsecn:factyuk}
Let us consider first the situation in which the torus is factorized
and the fluxes (or the angles) for all the branes involved in the
amplitude respect the factorization so that the reflection matrices,
and hence the monodromy matrices $R$ for the various open strings
commute. In this case there is a single complex basis of bosonic and
fermionic world-sheet fields, $\mathcal{Z}^i$ and $\Psi^i$, in which
all the monodromies act diagonally as specified by their eigenvalues
$\theta_i$, $\nu_i$ and $\omega_i$. We can then directly substitute
into the amplitude $\mathcal{A}(\theta,\nu,\omega)$ the expressions of
the vertices given in \secn{secn:spectrum}.

Both the NS vertex $V_{\phi^1_\theta}$ given in \Eq{vertexphi} and the
R vertices $V_{\lambda_\nu}$ and $V_{\lambda_\omega}$ (given by
\Eq{vertexR} with, obviously, the $\theta$'s replaced by $\nu$'s or
$\omega$'s) contain\footnote{They also contain superghost terms and
  $\ex{\ii k\cdot X}$ terms, but it is easy to see that their
  correlators do not modify the fusion coefficients, a part from
  imposing the obvious momentum conservation.} product of bosonic
twist fields $\sigma_{\theta_i}$ and of fermionic ones.

From the fermionic twist fields we get the correlators
\begin{equation}
\label{yuk6}
\begin{aligned}
{}&\langle\mathcal{S}_{\theta_1 - 1}(x_1) \mathcal{S}_{\nu_1-\frac 12}(x_2)
\mathcal{S}_{\omega_1 -\frac 12}(x_3)\rangle
\times
\langle\mathcal{S}_{\theta_2}(x_1) \mathcal{S}_{\nu_2-\frac 12}(x_2)
\mathcal{S}_{\omega_2-\frac 12}(x_3)\rangle
\\
{}& \times
\langle\mathcal{S}_{\theta_3}(x_1) \mathcal{S}_{\nu_3-\frac 12}(x_2)
\mathcal{S}_{\omega_3- \frac 12}(x_3)\rangle~~.
\end{aligned}
\end{equation}
Using, for instance, the bosonized formalism introduced in
\Eq{bosonization} it is immediate to see that the non-vanishing of
these correlators requires
\begin{equation}
\label{yuk7}
\begin{aligned}
\theta_1 + \nu_1 + \omega_1 & = 2~~,
\\
\theta_2 + \nu_2 + \omega_2 & = 1~~,
\\
\theta_3 + \nu_3 + \omega_3 & = 1~~.
\end{aligned}
\end{equation}
Subtracting the first of these equations from the sum of the others
yields a relation between the masses of the scalar components
$\phi^1_{\theta,\nu,\omega}$ of the three multiplets involved in the
interaction: using \Eq{scalars2} we find indeed
\begin{equation}
\label{yuk8}
M^2(\theta) + M^2(\nu) + M^2(\omega) = 0~~.
\end{equation}
This relation is obviously satisfied in supersymmetric configurations
of all the three open strings, i.e. when $\theta_2 + \theta_3 -
\theta_1 = 0$ and similarly for the $\nu's$ and the $\omega$'s so that
all scalar masses vanishes. If we allow for non-zero masses as
explained in~(\ref{thetaalpha}-\ref{mass2}), then this relation
implies that at least one of the three multiplets has a scalar
component which is tachyonic. In fact, this might be a desirable
feature: following the idea of the Higgs as a tachyon
\cite{Cremades:2002cs}, the Yukawa couplings correctly represent the
3-point functions among the Higgs field and the SM fermions.

The correlator of our vertices involves also, and this is in fact the
most crucial and non-trivial ingredient, the following correlator of
the bosonic twist fields
\begin{equation}
\label{yuk9}
\langle\sigma_{\theta_1}(x_1) \sigma_{\nu_1}(x_2) \sigma_{\omega_1}(x_3)\rangle
\times
\langle\sigma_{\theta_2}(x_1) \sigma_{\nu_2}(x_2) \sigma_{\omega_2}(x_3)\rangle
\times
\langle\sigma_{\theta_3}(x_1) \sigma_{\nu_3}(x_2) \sigma_{\omega_3}(x_3)\rangle~~.
\end{equation}
As for the fermionic twist fields, we get the product of three
independent correlators, pertaining to the CFT's of the three bosons
$\mathcal{Z}^i$, because of the factorized situation we are
considering.

For the CFT of a complex boson, the correlator of three twist fields
can be obtained from factorization of the 4-twist correlator and it
turns out~\cite{Dixon:1986qv,Burwick:1990tu,Erler:1992gt,Stieberger:1992bj,Stieberger:1992vb}
to be given by
\begin{equation}
\label{yuk10}
\langle\sigma_{\theta}(x_1) \sigma_{\nu}(x_2) \sigma_{\omega}(x_3)\rangle
\!=\!
\prod_{i>j}(x_i - x_j)^{h - 2 (h_i + h_j)} \times\!
\left\{
\begin{array}{lr}
\!\!\!\left(\frac{\Gamma(1 - \theta)}{\Gamma(\theta)}\frac{\Gamma(1 - \nu)}{\Gamma(\nu)}
\frac{\Gamma(1 - \omega)}{\Gamma(\omega)}\right)^{\frac 14}\, , & \theta + \nu + \omega =  1\\
\!\!\!\left(\frac{\Gamma(\theta)}{\Gamma(1 - \theta)}\frac{\Gamma(\nu)}{\Gamma(1 - \nu)}
\frac{\Gamma(\omega)}{\Gamma(1 - \omega)}\right) ^{\frac 14}\, ,&  \theta + \nu + \omega =  2
\end{array}
\right.
\end{equation}
where $h_i \equiv h_{\sigma_{\theta^i}}$ is the conformal dimension of
the twist field given in \Eq{hsigma}, and $h = \sum_i h_i$.  Inserting
this result in the product of bosonic twist correlators~\eq{yuk9},
upon taking into account the relations~\eq{yuk7} among the twist
angles, we finally can write the fusion coefficient of the three
vertices $V_{\phi^1_\theta}$, $V_{\lambda_\nu}$ and
$V_{\lambda_\omega}$
\begin{equation}
\label{yuk11}
\begin{aligned}
\mathcal{A}(\theta,\nu,\omega)
& = \! \left(
\frac{\Gamma(\theta_1)}{\Gamma(1 - \theta_1)} \prod_{i\not=1}\frac{\Gamma(1 -
\theta_i)}{\Gamma(\theta_i)}
\,
\frac{\Gamma(\nu_1)}{1 - \Gamma(\nu_1)}\prod_{i\not=1}\frac{\Gamma(1 -
\nu_i)}{\Gamma(\nu_i)}
\,
\frac{\Gamma(\omega_1)}{\Gamma(1 -
\omega_1)}\prod_{i\not=1}\frac{\Gamma(1 - \omega_i)}{\Gamma(\omega_i)}
\right)^{\frac 14}
\\
& =
\Big[K(\theta) K(\nu) K(\omega)\Big]^{-\frac 12}
\end{aligned}
\end{equation}
where in the last line we used the explicit expression of the K\"ahler
metrics for the chiral multiplets, given in \Eq{gkahlers} (see also
\Eq{gkahler} in the field theory limit). Notice that the exponential
terms in $K(\theta) K(\nu) K(\omega)$ reduce to 1 using the relation
\Eq{yuk8} between the masses. Thus, the explicit computation of the
quantum part of the stringy Yukawa couplings in the case of factorized
twisted chiral multiplets agrees with the expectation \Eq{yuk4}
inferred from the non-renormalization property of the superpotential.
Of course, the same property can be derived also by working with
the symmetric scalar, following the reasoning for the K\"ahler metric
before Eq.~(\ref{finmetsym}), finding agreement.

\vskip 0.5cm
\subsection{The oblique case and non-abelian twists fields}
\label{subsecn:nonfactyuk}
\begin{figure}[htp]
\hspace{.5cm}
\begin{center}
\begin{picture}(0,0)%
\includegraphics{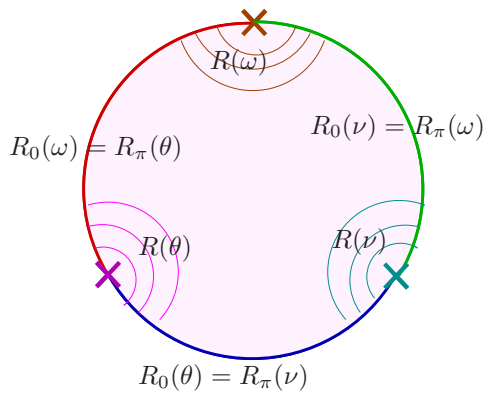}%
\end{picture}%
\input yuk_3.pstex_t
\end{center}
\caption{\small The reflection matrices $R_0$ and $R_\pi$ and the
monodromy matrices $R = R_\pi^{-1} R_0$ for the three twisted open
strings. The matrices pertaining to the different strings are labeled
by the angles $\theta$, $\nu$, $\omega$ which determine the
eigenvalues of the monodromy according to \Eq{diagonalR1}.}
\label{fig:yukbis}
\end{figure}

Let us now suppose that the monodromy matrices pertaining to the three
open strings we are considering do not commute with each other. This
is what happens for generic quantized fluxes on each stack of branes
(i.e., for ``oblique fluxes'') on a generic torus. Indeed the set of
reflection matrices for strings with their endpoints $\sigma=0,\pi$
attached to D-branes with fluxes $F_\sigma$, given by \Eq{R}, have no
reason to commute between themselves, except in the factorized case
considered above.  Hence, the various monodromy matrices $R =
R_\pi^{-1} R_0$ do not commute either.
Each monodromy
can still be diagonalized as in \Eq{diagonalR1}, and its eigenvalues depend on a
set of angles $\theta_i$.
However, the monodromies $R(\theta)$, $R(\nu)$ and $R(\omega)$
of the three open strings involved in the Yukawa amplitude cannot be
\emph{simultaneously} diagonalized\footnote{They are not completely
independent, though, since $R(\theta)R(\nu)R(\omega)=\mathbf{1}$, see
Figure \ref{fig:yukbis}.}.

The bosonic and fermionic twist fields occurring in a vertex must be such that their
OPE with the bosonic
(resp. fermionic) fields impose the condition
\begin{equation}
\label{yuk12}
X^M(\ex{2\pi\ii}z) = \left[R(\theta)\right]^M_{~N}X^N(z)
\end{equation}
for the six bosonic coordinates $X^M$ along the torus (resp., the
analogous relation for the $\Psi^M$ fields) as indicated in \Eq{chiraly}.  These
fields can therefore be defined only within the CFT describing the six
bosonic (resp. fermionic) directions along $\mathcal{T}_6$, and not
within its factorization in the CFT's of three complex bosons (resp.
fermions).  We use the notation $\hat{\sigma}_{R(\theta)}$ for such
bosonic twist fields and $\hat{\mathcal{S}}_{R(\theta)}$ for the
fermionic ones.

According to Eqs.~(\ref{yuk5.2}) and (\ref{yuk5.1}), the quantum
Yukawa amplitude\footnote{The notation here is slightly
misleading. The angles $\theta,\nu,\omega$ just label the type of
vertices in the amplitude.  There is no reason a priori, from the CFT
point of view, that the amplitude actually be a function of these
angles only. We would, in general, expect that it depends of the
complete monodromy matrices $R(\theta)$, $R(\nu)$, $R(\omega)$.}
$\mathcal{A}(\theta,\nu,\omega)$ coincides with the fusion coefficient
in the correlator of the three emission vertices $V_{\phi^1_\theta}$,
$V_{\lambda_\nu}$ and $V_{\lambda_\omega}$. It is thus determined by
the product of the three-point correlators of the bosonic and
fermionic twist fields appearing in these vertices
\begin{equation}
\label{yuk13}
\langle \hat{\mathcal{S}}^\prime_{R(\theta)}(x_1)
\hat{\mathcal{S}}^{\mathrm{Ramond}}_{R(\nu)}(x_2)
\hat{\mathcal{S}}^{\mathrm{Ramond}}_{R(\omega)}(x_3)\rangle
\times
\langle \hat{\sigma}_{R(\theta)}(x_1) \hat{\sigma}_{R(\nu)}(x_2)
\hat{\sigma}_{R(\omega)}(x_3)\rangle~~.
\end{equation}
Here we denoted as $\hat{\mathcal{S}}^\prime_{R(\theta)}$ the
excited\footnote{Of course, we can diagonalize \emph{one} of the
  monodromy matrices, say $R(\theta)$ by choosing its complex
  eigenvector basis $\mathcal{Z}$. The corresponding open string
  sector is then described exactly as in Section
  \ref{secn:modeexpansions}. The twist fields are factorized:
  $\hat\sigma_{R(\theta)}=
  \sigma_{\theta_1}\sigma_{\theta_2}\sigma_{\theta_3}$, and similarly
  for the fermionic ones.  The vertex $V_{\phi^1_\theta}$ has the
  expression of \Eq{vertexphi}, and the excited fermionic twist
  $\hat{\mathcal{S}}^\prime_{R(\theta)}$ it contains is just
  $\mathcal{S}_{\theta_1-1}\mathcal{S}_{\theta_2}\mathcal{S}_{\theta_3}$.
}  fermionic twist field which enters the emission vertex
$V_{\phi^1_\theta}$, and by
$\hat{\mathcal{S}}^{\mathrm{Ramond}}_{R(\nu)}$ the fermionic twists in
the Ramond sector, which implement an extra minus sign in the
monodromy with respect to the NS ones.
If the three monodromy matrices $R(\theta)$, $R(\nu)$ and
$R(\omega)$ commute with each other, we can diagonalize
simultaneously the twist operators $\hat \sigma_R(\theta)$,
$\hat \sigma_R(\nu)$ and $\hat \sigma_R(\omega)$, and thus the
last correlator in (\ref{yuk13}) can be factorized into a product
of three correlators as in (\ref{yuk9}). Notice that this may
happen also for a non-diagonal background. However, in the most
general situation the three monodromy matrices do not
commute and the structure of the bosonic twist correlator is more
involved.

On the other hand, the non-renormalization property of the
superpotential $W$ still suggests that the amplitude
$\mathcal{A}(\theta,\nu,\omega)$ should be given in terms of the
K\"ahler metrics for the three chiral multiplets, as in \Eq{yuk4}
\begin{equation}
\label{yuk14}
\mathcal{A}(\theta,\nu,\omega) = \Big[K(\theta) K(\nu) K(\omega)\Big]^{-\frac 12}~~.
\end{equation}
We have shown in this paper that the expression of the K\"ahler metric
is always given by \Eq{gkahlers}, independently of whether we are in an
abelian or in a oblique situation, and depends just on the monodromy
eigenvalues. So, in fact, $\mathcal{A}(\theta,\nu,\omega)$ should
really depend just on the angles $\theta_i$, $\nu_i$ and $\omega_i$.

As a consequence, we are lead to conjecture that the
\emph{non-abelian} twist field correlator in \Eq{yuk13}, which in
principle depends on the entire monodromy matrices $R(\theta)$,
$R(\nu)$ and $R(\omega)$, has in fact the same expression of a
correlator of abelian twist fields characterized by the monodromy
eigenvalues $\theta_i$, $\nu_i$, $\omega_i$.
Proving (or disproving) this conjecture is a very interesting
challenge in CFT.

\vskip 1cm
\noindent {\large {\bf Acknowledgments}} \vskip 0.2cm
\noindent We thank Laurent Gallot for collaboration at the
beginning of this project. We also thank Bobby Acharya, Massimo Bianchi, Giulio Bonelli,
Gianguido Dall'Agata, Dario Du\`o, Marialuisa Frau, Wolfang
Lerche, Igor Pesando, Claudio Scrucca, Marco Serone, Gary Shiu,
Stephan Stieberger and Angel Uranga for useful discussions and comments. This work is
partially supported by the European Community's Human Potential
Programme under contract MRTN-CT-2004-005104 (in which A.L. is
associated to Torino University and M.Be. to Padova University)
and by the Italian MIUR under contract PRIN-2003023852. M.Be. is
also supported by a MIUR fellowship within the program
``Incentivazione alla mobilit\`a degli studiosi italiani e
stranieri residenti all'estero''.

\vspace{0.25cm}
\appendix

\sect{Dependence of the open string twists from the closed moduli}
\label{app:jac}

Let $m$ be a closed string modulus which is a generic function of the
NS-NS fields $G$ and $B$. {F}rom~\Eq{diagonalR1} we have
\begin{equation}
\label{dmt1}
\begin{aligned}
2\pi\ii \frac{\partial\theta_i}{\partial m} & = \left(\frac{\partial \RCB}{\partial m}
\RCB^{-1}\right)_{ii}
= \left(\ECB \frac{\partial \RB}{\partial m} \RB^{-1} \ECB^{-1} +
\comm{\frac{\partial \ECB}{\partial m}\ECB^{-1}}{\RCB} \RCB^{-1}
\right)_{ii}
\\
& = \left(\ECB \frac{\partial \RB}{\partial m} \RB^{-1} \ECB^{-1}\right)_{ii}~~,
\end{aligned}
\end{equation}
where in the last step we used the fact that the commutator of an
arbitrary matrix with a diagonal matrix, such as $\RCB$, has no
entries on the diagonal.  Since $\RB = \RB_\pi^{-1}\RB_0$, we get with
simple manipulations
\begin{equation}
\label{dmt2}
2\pi\ii \frac{\partial\theta_i}{\partial m} =
\left(\ECB
\left[\frac{\partial \RB_\pi^{-1}}{\partial m} \RB_\pi -
\RB \frac{\partial \RB_0^{-1}}{\partial m} \RB_\pi\right]
\ECB^{-1}\right)_{ii}~~.
\end{equation}
We can now take advantage of the following property (which will be
needed again in later stages of the computation):
\begin{equation}
\label{dmt3}
\left(\ECB \RB \AB \ECB^{-1}\right)_{ii} =
\left(\ECB  \AB \RB \ECB^{-1}\right)_{ii}~~,
\end{equation}
which holds for any matrix $\AB$, since $\ECB\RB\ECB^{-1} = \RCB$ is
diagonal, to get
\begin{equation}
\label{dmt4}
2\pi\ii \frac{\partial\theta_i}{\partial m} =
\left(\ECB
\left[\frac{\partial \RB_\pi^{-1}}{\partial m} \RB_\pi -
\frac{\partial \RB_0^{-1}}{\partial m} \RB_0\right]
\ECB^{-1}\right)_{ii}~~.
\end{equation}
{F}rom the expression (\ref{R}) of the reflection matrices $\RB_\sigma$
($\sigma=0,\pi$) it follows that
\begin{equation}
\label{dmt5}
\frac{\partial \RB_\sigma^{-1}}{\partial m} \RB_\sigma
=  \frac 12 \left(\mathbf{1} + \RB_\sigma^{-1}\right) \GB^{-1}
\left(-\frac{\partial(\GB + \BB)}{\partial m} +
\frac{\partial(\GB - \BB)}{\partial m} \RB_\sigma
\right)~~,
\end{equation}
where we used
\begin{equation}
(\GB + \BB + \FB_\sigma)^{-1} = \frac 12 \left(\mathbf{1} +
  \RB_\sigma^{-1}\right) \GB^{-1}~~.
\label{gtr}
\end{equation}
Substituting the expression~(\ref{dmt5}) into~Eq. (\ref{dmt4}) we get four contributions.
Combining the two terms proportional to $ \partial(\GB + \BB)$ we find
\begin{equation}
\label{dmt6}
-\frac 12 \left(\ECB
\left[\RB_\pi^{-1} - \RB_0^{-1}\right]\GB^{-1} \frac{\partial(\GB + \BB)}{\partial m}
\ECB^{-1}\right)_{ii}~~.
\end{equation}
Again, by using~Eq. (\ref{gtr}), the two terms proportional to $ \partial(\GB
- \BB)$ can be written as
\begin{eqnarray}
\label{dmt7}
&&  \frac 12 \left(\ECB
\left[
  (\mathbf{1}+\RB_\pi^{-1}) \GB^{-1} \frac{\partial(\GB - \BB)}{\partial
  m}  R_0 R_0^{-1}  \RB_\pi - (\mathbf{1}+\RB_0^{-1}) \GB^{-1}
  \frac{\partial(\GB - \BB)}{\partial m} \RB_0 \right]
 \ECB^{-1}\right)_{ii}
\nonumber
\\
&= &
\frac 12 \left(\ECB \left[\RB_0^{-1} \RB_\pi -\mathbf{1} \right]
\GB^{-1} \frac{\partial(\GB - \BB)}{\partial m}
\RB_0\ECB^{-1}\right)_{ii}
\\
&=&
\frac 12 \left(\ECB
 \GB^{-1} \frac{\partial(\GB - \BB)}{\partial m} \left[\RB_\pi -
   \RB_0\right]
\ECB^{-1}\right)_{ii}
\nonumber
\end{eqnarray}
where we used several times the property (\ref{dmt3}) to move the matrix
$\RB^{-1}$. The last step is to write Eq.~(\ref{dmt6}) in terms of the
transpose matrix. This can be done by using the properties (\ref{Gcal}) of
the vielbeins ${}^t {\cal E} \GCB = G {\cal E}^{-1}$ and of the
reflection matrix (\ref{orthogonality}). In this way we find that
Eq.~(\ref{dmt6}) can be written in terms of the anti-holomorphic elements
of the same matrix appearing in the last step of Eq.~(\ref{dmt7})
\begin{equation}
\label{dmt8again}
 2\pi\ii \frac{\partial\theta_i}{\partial m} =
\frac 12 \left[\ECB
 \GB^{-1} \frac{\partial(\GB - \BB)}{\partial m} \left[\RB_\pi - \RB_0\right]
\ECB^{-1}\right]_{ii} -
\frac 12 \left[\ECB
 \GB^{-1} \frac{\partial(\GB - \BB)}{\partial m} \left[\RB_\pi - \RB_0\right]
\ECB^{-1}\right]_{\overline i\, \overline i}~~.
\end{equation}
If $m$ is a real modulus, then the second term in the equation above
 is just the complex conjugate of the first one.

\vskip 0.5cm
\subsection{The two-dimensional case}
\label{subapp:jac2} Let us check the general expression of the
dependence of the open string twists from the closed string moduli
in the simple case of a 2-dimensional torus. From the results of
Section~\ref{secn:shifts} we know that the twists $\theta$ depend only
on $T$
\begin{equation}
\label{dertdir}
2\pi\ii\frac{\partial\theta}{\partial U} = 0~~,~~~~~
 2\pi\ii\frac{\partial\theta}{\partial T}  =
\frac{ f_\pi - f_0}{(T - f_\pi)(T -
f_0)}~~.
\end{equation}

Let us retrieve the same result from Eq.~(\ref{dmt8again}). Since in the
present case the reflection matrices $\RCB_\sigma$ in the complex
basis are diagonal, it is convenient to rewrite Eq.~(\ref{dmt8again}) by suitably
inserting the identity written as ${}^t \ECB\,\, {}^t\ECB^{-1}$ or
$\ECB\ECB^{-1}$, getting
\begin{equation}
\label{dertjac2}
\begin{aligned}
2\pi\ii\frac{\partial\theta}{\partial m} & =
\frac 12  \left(
{}^t\ECB^{-1} \frac{\partial (\GB - \BB)}{\partial m} \ECB^{-1}
\right)_{21} (\RCB_\pi - \RCB_0)_{11}
\\
& - \frac 12  \left(
{}^t\ECB^{-1} \frac{\partial (\GB - \BB)}{\partial m} \ECB^{-1}
\right)_{12} (\RCB_\pi - \RCB_0)_{22}
~~.
\end{aligned}
\end{equation}
By using (\ref{GB2}) and (\ref{E2d}), we find for $m=U$
\begin{equation}
\label{chkU2}
{}^t\ECB^{-1} \frac{\partial (\GB \pm \BB)}{\partial U} \ECB^{-1} =
\frac{\ii}{U_2}\begin{pmatrix}
0 & 0 \\ 0 & 1
\end{pmatrix}~~.
\end{equation}
Since this matrix has no $(21)$ component, we get immediately from
\Eq{dertjac2} that $\partial\theta/\partial U = 0$, in agreement with
\Eq{dertdir}. For $m=T$, we get instead
\begin{equation}
\label{chkT2}
{}^t\ECB^{-1} \frac{\partial (\GB - \BB)}{\partial T} \ECB^{-1} =
-\frac{\ii}{T_2}
\begin{pmatrix}
0 & 0 \\ 1 & 0
\end{pmatrix}~~.
\end{equation}
This matrix has a non-vanishing $(21)$ component and so contributes to
the first term of \Eq{dertjac2} and we get
\begin{equation}
\label{chkT22}
\begin{aligned}
2\pi\ii\frac{\partial\theta}{\partial T}
& =  \frac{-\ii}{2T_2}
(\RCB_\pi - \RCB_0)_{11} =
\frac{\ii}{2T_2} \left(\frac{\bar T -  f_\pi}{T -  f_\pi}
- \frac{\bar T -  f_0}{T -  f_0}\right)
\\ & = \frac{ f_\pi -  f_0}{(T -  f_\pi)(T -  f_0)}~~,
\end{aligned}
\end{equation}
in perfect agreement with \Eq{dertdir}.

\vskip 0.5cm
\sect{The metric for the untwisted matter}
\label{app:unt}

In this Appendix we apply the same technique described in
Section~\ref{secn:kahler} to the case of open strings starting and
ending on the same stack of D9-branes. In particular, we show that it
is possible to determine completely the metric of the untwisted scalars from a
3-point function involving two scalars and one closed string
modulus. Of course, this result can be read directly by
compactifying the Born-Infeld action of a D9-brane to four
dimensions
\begin {equation}
\label{unac}
S_{\rm untw} =-\frac{\cal N}{2} \int d^4x\, {\rm e}^{-\phi_4}
\frac{\sqrt{{\rm det}(G-{\cal F})}}{({\rm det}\,G)^{{1}/{4}}}
\,D^\mu A_M D_\mu A_N\, G_{\rm open}^{MN}
\end{equation}
where ${\cal N}=T_9\,(2\pi\alpha')^{3/2}$, with $T_9$ being the
D9-brane tension, and the four dimensional dilaton $\phi_4$ is
related to the ten dimensional one by
${\rm e}^{-\phi_4}={\rm e}^{-\phi_{10}}\,({\rm
det}\,G)^{{1}/{4}}\,(2\pi\alpha')^{-3/2}$ (recall that in our
conventions the internal metric $G_{MN}$ has dimensions of (length)$^2$
and that the fields $A_M$ are dimensionless).
Notice the appearance in (\ref{unac}) of the open string metric~\cite{Seiberg:1999vs}
\begin{equation}
\label{openG}
G_{\rm open}^{MN}\equiv
\frac{1}{2} G^{MN}+\frac{1}{4} (R_0)^{MN}+\frac{1}{4}({}^t R_0)^{MN}=
\left[\frac{1}{G-{\cal F}}\right]^{(MN)}~~.
\end{equation}
Usually the action $S_{\rm untw}$ is written in terms of a rescaled string
coupling~\cite{Seiberg:1999vs}; here, instead, we prefer to keep the dependence
on the four dimensional dilaton $\phi_4$, because the insertion of the closed
string modulus $m$ gives a differential equation where $\phi_4$ is
kept constant, as seen in Section~\ref{secn:shifts}.

The dimensionless fields $A_M$ in~\eq{unac} simply represent the internal components of the gauge
field and so the corresponding vertex operator (in the zero picture)
can be read from the boundary terms of the actions~\eq{Sx}
and~\eq{Sferm}, namely
\begin{equation}
V_{\phi_M} = \phi_M(k) \Big( \frac{1}{2}\partial x^M+\frac{1}{2}\bar{\partial}
x^M - \frac{\ii}{2}
k_\mu \psi_-^\mu \,\psi_+^M
- \frac{\ii}{2}
k_\mu \psi_+^\mu \, \psi_-^M\Big)\,{\rm e}^{\ii\,k\cdot X}
\end{equation}
where $\phi_M=A_M/\sqrt{2\alpha'}$.
On the other hand, the vertex for a closed string modulus $m$
is given by~\Eq{vm}, where in the $(-1,-1)$ picture the operator
$W^{MN}(z,\overline z)$ is the product of the left and right
vertices of \Eq{vm-1}.

Now we can proceed as in Section~\ref{secn:kahler} and compute the
amplitude
\begin{equation}
\label{ampdef0}
{\cal A}_{\rm untw} =
\frac{{\rm e}^{-\ii\pi\alpha' k_L \cdot k_R}}{8\pi {\alpha'}^2}\,
\int \frac{dx_1 \,dx_2 \,d^2z}{dV_{\rm{CKG}}}\,
\big\langle V_{\phi_M}(x_1)\, W_m(z,\overline z) \,V_{\phi_N}(x_2)\big\rangle
\end{equation}
where in the normalization we have included also the cocycle factor of the closed string
vertex. The boundary conditions for this diagram are
\begin{equation}
\psi_\pm^\mu(x)= \psi^\mu(x)~~~~,~~~~\psi_+^M(x)= \psi^M(x)~~~~,~~~~
\psi_-^M(x)=\big(R_0\big)^M_{~N}\,\psi^N(x)
\label{bcopen}
\end{equation}
for the fermionic open string coordinates, and
\begin{equation}
\psi_L^M(z)= \psi^M(z)~~~~,~~~~
\psi_R^M(\overline z)=\big(R_0\big)^M_{~N}\,\psi^N(\overline z)
\label{bcclosed}
\end{equation}
for the closed string ones. Then by following the same steps as in
Section~\ref{secn:kahler}, we obtain
\begin{eqnarray}
\label{amplun}
{\cal A}_{\rm untw}&=& \frac{{\rm e}^{-\ii\pi\alpha' k_L \cdot k_R}}{32\pi {\alpha'}^3}\,
A_M(k_1) \,A_N(k_2)\,\left(\frac{\partial}{\partial m}(G-B)\, R_0\right)_{PQ}
~\int \frac{dx_1 \,dx_2 \,d^2z}{dV_{\rm{CKG}}}
~~ {\cal Y}
\nonumber \\ && \times~(z-\bar{z})^{-1}
\Big(\big\langle \partial x^M(x_1) \, \partial x^N(x_2) \big\rangle
~\big\langle \psi^P(z)\,\psi^Q(\overline z)\big\rangle \\
&&+\,\frac{\alpha'\,k_{1}\cdot k_{2}}{2}\,(x_1-x_2)^{-1} (1+R_0)^M_{~I} (1+R_0)^N_{~J}\big\langle
\psi^I (x_1)\, \psi^P(z)\,\psi^Q(\overline z)\,\psi^J(x_2)
\big\rangle \Big)\nonumber
\end{eqnarray}
where
\begin{equation}
{\cal Y}\equiv
\big\langle \ex{\ii k_1\cdot X(x_1)}\,\ex{\ii k_L \cdot X(z)} \,\ex{\ii k_R\cdot
 X(\bar{z})}\,
 \ex{\ii k_2 \cdot X(x_2)} \big\rangle =
 \omega^{-\alpha's/2}\,(1-\omega)^{\alpha's}
\end{equation}
in terms of the anharmonic ratio $\omega$ defined in \Eq{wdef}.
To proceed, we use the following basic correlators
\begin{equation}
\big\langle x^M(x_1) x^N(x_2) \big\rangle= -2 \alpha' \, G_{\rm open}^{MN}  \ln   (x_1-x_2)
~~~~\mbox{and}~~~~
\big\langle \psi^M(x_1) \, \psi^N (x_2) \big\rangle=
\frac{2\a' G^{MN}}{(x_1-x_2)}
\label{bas0}
\end{equation}
and Eq. (\ref{measure}), and, after simple manipulations, find
\begin{eqnarray}
{\cal A}_{\rm untw}&=&
\frac{{\rm e}^{-\ii\,\pi\alpha' s/2}}{8\pi {\alpha'}}\,
A_M(k_1) \,A_N(k_2)\,\left(\frac{\partial}{\partial m}(G-B)\, R_0\right)_{PQ}
~\int_{\mathcal{C}} d\omega~ \omega^{-\alpha' s/2}(1-\omega)^{\alpha' s-2}\,
\nn\\
&& \Big[
- G_{\rm open}^{MN}\, G^{PQ}
(1-\a's)+\frac{\a's}{4}\,\big(G^{PM}+{}^t R_0^{PM}\big) \big(G^{QN}+{}^t R_0^{QN}\big)\,
\frac{1-\omega}{\omega}
\nn\\
&&~~-
\frac{\a's}{4} (G^{QM}+{}^t R_0^{QM}) (G^{PN}+{}^t R_0^{PN})(1-\omega)
  \Big]~~.
\end{eqnarray}
In this expression, the last two terms (proportional to $\alpha's$) are one the transpose
of the other, as one can see with the change of variable $\omega \to
1/\omega$; so we can keep just the first term and symmetrize the
result in the indices $M$ and $N$. In this way we get
\bea
{\cal A}_{\rm untw}&=&  \frac{\ii}{4\pi\alpha'}\, A_M(k_1)
\,A_N(k_2) \left(\frac{\partial}{\partial m}(G-B)\, R_0\right)_{PQ}
\sin(\pi \alpha' s/2)\nonumber \\
&&\Big[-(1-{\a's})\,B(1-{\alpha's}/{2},\alpha' s-1)\,
G_{\rm open}^{MN}\, G^{PQ} \\
&&~~+
\frac{\a's}{2}\, B(-\alpha's/2,\alpha' s)
\, \big(G^{P(M}+{}^t R_0^{P(M}\big)\, \big(G^{QN)}+{}^t R_0^{QN)}\big)\Big]~~.
\nonumber
\eea
Now we expand this result in $\alpha's$ and focus on the (leading) term with
two derivatives which captures the kinetic term for the scalar
fields $A_M$, that is
\begin{equation}
\begin{aligned}
{\cal A}_{\rm untw} =&~\frac{\ii}{2}\,k_1\cdot k_2\, A_M(k_1) \,A_N(k_2) \Bigg\{
\frac{1}{4}\,{\rm tr}\Big[\frac{\partial}{\partial m}(G-B)\,
R_0\,G^{-1}\Big]
G_{\rm open}^{MN}
\\
&-\Big[(G-{\cal F})^{-1} \frac{\partial}{\partial m}(G-B)\, R_0
(G+{\cal F})^{-1}\Big]^{(MN)}\,\Bigg\}
\end{aligned}
\label{resultunt}
\end{equation}
where we have used ${}^t R_0 = G R_0^{-1} G^{-1}$ and the
identity~\eq{gtr}
to rewrite the last term.
The term proportional to $G_{\rm open}^{MN}$ yields the same total derivative
found in Section~3 of~\cite{Lust:2004cx}, namely
\begin{equation*}
\frac{1}{4}\, {\rm tr}\Big[\frac{\partial}{\partial m}(G-B)\, R_0\,G^{-1}\Big]=
\frac{1}4 \,{\rm tr}\Big[\frac{\partial}{\partial m}(G-{\cal F})
\Big(2(G-{\cal F})^{-1}-G^{-1}\Big)\Big]
=\partial_m \ln K
\end{equation*}
where \footnote{Here, we have included also the dependence on the dilaton $\phi_4$
which could be fixed by computing a 3-point amplitude with the dilaton
vertex, as discussed in detail in Ref. \cite{Lust:2004cx}. We have also included the
appropriate dimensional prefactor ${\cal N}=T_9\,(2\pi\alpha')^{3/2}$ to make $K$ dimensionless.}
\begin{equation}
K = {\cal N}\,{\rm e}^{-\phi_4}\,\frac{\sqrt{{\rm det}(G-{\cal F})}}{({\rm det}G)^{1/4}}~~.
\end{equation}
The second term in~\eq{resultunt} yields exactly the open string
metric~\eq{openG}; indeed
\begin{equation}
\Big[(G-{\cal F})^{-1} \frac{\partial}{\partial m}(G-B)\, R_0
(G+{\cal F})^{-1}\Big]^{(MN)}=-\frac{\partial}{\partial m}
\Big[\frac{1}{G-{\cal F}}\Big]^{(MN)}=-\frac{\partial}{\partial m} G_{\rm open}^{MN}~~.
\end{equation}
Thus, we can write our result as
\begin{equation}
{\cal A}_{\rm untw} = \ii\,K^{-1}\,\frac{\partial}{\partial m}
\Big[\frac12\,K\,G_{\rm
open}^{MN}\,\,k_1\cdot k_2\,A_M(k_1)\,A_n(k_2)\Big]~~.
\label{finale}
\end{equation}
In complete analogy with what we did in Eq. (\ref{lagder}) for the twisted scalars,
we identify the square bracket of Eq. (\ref{finale}) with the
(momentum space) Lagrangian of the untwisted fields
and thus reconstruct the action~\eq{unac}, in perfect agreement with the Born-Infeld result.


\begin{thebibliography}{10}

\bibitem{Bachas:1995ik}
C.~Bachas,
\newblock (1995), hep-th/9503030.

\bibitem{Berkooz:1996km}
M.~Berkooz, M.~R. Douglas, and R.~G. Leigh,
\newblock Nucl. Phys. {\bf B480}, 265 (1996), hep-th/9606139.

\bibitem{Rabadan:2001mt}
R.~Rabadan,
\newblock Nucl. Phys. {\bf B620}, 152 (2002), hep-th/0107036.

\bibitem{Blumenhagen:1999ev}
R.~Blumenhagen, L.~Gorlich and B.~Kors,
\newblock  JHEP {\bf 0001} (2000) 040, hep-th/9912204.

\bibitem{Blumenhagen:2000wh}
R.~Blumenhagen, L.~Goerlich, B.~Kors and D.~Lust,
\newblock  JHEP {\bf 0010} (2000) 006, hep-th/0007024.

\bibitem{Angelantonj:2000hi}
C.~Angelantonj, I.~Antoniadis, E.~Dudas and A.~Sagnotti,
\newblock  Phys.\ Lett.\ B {\bf 489} (2000) 223, hep-th/0007090.

\bibitem{Aldazabal:2000cn}
G.~Aldazabal, S.~Franco, L.~E. Ibanez, R.~Rabadan, and A.~M. Uranga,
\newblock JHEP {\bf 02}, 047 (2001), hep-ph/0011132.

\bibitem{Uranga:2003pz}
A.~M. Uranga,
\newblock Class. Quant. Grav. {\bf 20}, S373 (2003), hep-th/0301032.

\bibitem{Kiritsis:2003mc}
E.~Kiritsis,
\newblock Fortsch. Phys. {\bf 52}, 200 (2004), hep-th/0310001.

\bibitem{Lust:2004ks}
D.~Lust,
\newblock Class. Quant. Grav. {\bf 21}, S1399 (2004), hep-th/0401156.

\bibitem{Kokorelis:2004tb}
C.~Kokorelis,
\newblock (2004), hep-th/0410134.

\bibitem{Blumenhagen:2005mu}
R.~Blumenhagen, M.~Cvetic, P.~Langacker, and G.~Shiu,
\newblock (2005), hep-th/0502005.

\bibitem{MarchesanoBuznego:2003hp}
F.~G. Marchesano~Buznego,
\newblock (2003), hep-th/0307252.

\bibitem{Ott:2003yv}
T.~Ott,
\newblock Fortsch. Phys. {\bf 52}, 28 (2004), hep-th/0309107.

\bibitem{Gorlich:2004zs}
L.~Gorlich,
\newblock (2004), hep-th/0401040.

\bibitem{Anastasopoulos:2005ba}
P.~Anastasopoulos,
\newblock (2005), hep-th/0503055.

\bibitem{Cremades:2002cs}
D.~Cremades, L.~E. Ibanez, and F.~Marchesano,
\newblock JHEP {\bf 07}, 022 (2002), hep-th/0203160.

\bibitem{Cremades:2003qj}
D.~Cremades, L.~E. Ibanez, and F.~Marchesano,
\newblock JHEP {\bf 07}, 038 (2003), hep-th/0302105.

\bibitem{Cvetic:2003ch}
M.~Cvetic and I.~Papadimitriou,
\newblock Phys. Rev. {\bf D68}, 046001 (2003), hep-th/0303083.

\bibitem{Abel:2003vv}
S.~A. Abel and A.~W. Owen,
\newblock Nucl. Phys. {\bf B663}, 197 (2003), hep-th/0303124.

\bibitem{Lust:2004cx}
D.~Lust, P.~Mayr, R.~Richter, and S.~Stieberger,
\newblock Nucl. Phys. {\bf B696}, 205 (2004), hep-th/0404134.

\bibitem{Cremades:2004wa}
D.~Cremades, L.~E. Ibanez, and F.~Marchesano,
\newblock JHEP {\bf 05}, 079 (2004), hep-th/0404229.

\bibitem{Abel:2003fk}
S.~A. Abel, M.~Masip, and J.~Santiago,
\newblock JHEP {\bf 04}, 057 (2003), hep-ph/0303087.

\bibitem{Abel:2003yh}
S.~A. Abel, O.~Lebedev, and J.~Santiago,
\newblock Nucl. Phys. {\bf B696}, 141 (2004), hep-ph/0312157.

\bibitem{Lust:2003ky}
D.~Lust and S.~Stieberger,
\newblock (2003), hep-th/0302221.

\bibitem{Bianchi:2005sa}
M.~Bianchi and E.~Trevigne,
\newblock (2005), hep-th/0506080.

\bibitem{Klebanov:2003my}
I.~R. Klebanov and E.~Witten,
\newblock Nucl. Phys. {\bf B664}, 3 (2003), hep-th/0304079.

\bibitem{Axenides:2003hs}
M.~Axenides, E.~Floratos and C.~Kokorelis,
\newblock  JHEP {\bf 0310} (2003) 006, hep-th/0307255.

\bibitem{Antoniadis:2004pp}
I.~Antoniadis and T.~Maillard,
\newblock Nucl. Phys. {\bf B716}, 3 (2005), hep-th/0412008.

\bibitem{Curio:2001qi}
G.~Curio and A.~Krause,
\newblock  Nucl.\ Phys.\ B {\bf 643} (2002) 131, hep-th/0108220.

\bibitem{Acharya:2002kv}
B.~S. Acharya,
\newblock (2002), hep-th/0212294.

\bibitem{Blumenhagen:2005tn}
R.~Blumenhagen, M.~Cvetic, F.~Marchesano, and G.~Shiu,
\newblock JHEP {\bf 03}, 050 (2005), hep-th/0502095.

\bibitem{Bianchi:2005yz}
M.~Bianchi and E.~Trevigne,
\newblock JHEP {\bf 08}, 034 (2005), hep-th/0502147.

\bibitem{Curio:2005ew}
G.~Curio, A.~Krause and D.~Lust,
\newblock (2005), hep-th/0502168.

\bibitem{Villadoro:2005cu}
G.~Villadoro and F.~Zwirner,
\newblock JHEP {\bf 06}, 047 (2005), hep-th/0503169.

\bibitem{DeWolfe:2005uu}
O.~DeWolfe, A.~Giryavets, S.~Kachru, and W.~Taylor,
\newblock JHEP {\bf 07}, 066 (2005), hep-th/0505160.

\bibitem{Antoniadis:2005nu}
I.~Antoniadis, A.~Kumar, and T.~Maillard,
\newblock (2005), hep-th/0505260.

\bibitem{Camara:2005dc}
P.~G. Camara, A.~Font, and L.~E. Ibanez,
\newblock JHEP {\bf 09}, 013 (2005), hep-th/0506066.

\bibitem{Lust:2005dy}
D.~Lust, S.~Reffert, W.~Schulgin, and S.~Stieberger,
\newblock (2005), hep-th/0506090.

\bibitem{GarciadelMoral:2005js}
M.~P. Garcia~del Moral,
\newblock (2005), hep-th/0506116.

\bibitem{Kumar:2005hf}
J.~Kumar and J.~D. Wells,
\newblock JHEP {\bf 09}, 067 (2005), hep-th/0506252.

\bibitem{Marchesano:2004yq}
F.~Marchesano and G.~Shiu,
\newblock  Phys.\ Rev.\ D {\bf 71} (2005) 011701, hep-th/0408059.

\bibitem{Ibanez:2004iv}
L.~E. Ibanez,
\newblock Phys. Rev. {\bf D71}, 055005 (2005), hep-ph/0408064.

\bibitem{Marchesano:2004xz}
F.~Marchesano and G.~Shiu,
\newblock JHEP {\bf 0411} (2004) 041, hep-th/0409132.

\bibitem{Cvetic:2005bn}
M.~Cvetic, T.~Li, and T.~Liu,
\newblock Phys. Rev. {\bf D71}, 106008 (2005), hep-th/0501041.

\bibitem{Dudas:2005jx}
E.~Dudas and C.~Timirgaziu,
\newblock Nucl. Phys. {\bf B716}, 65 (2005), hep-th/0502085.

\bibitem{Chen:2005mj}
C.-M. Chen, T.~Li, and D.~V. Nanopoulos,
\newblock (2005), hep-th/0509059.

\bibitem{Coriano':2005js}
C.~Corian\'o, N.~Irges, and E.~Kiritsis,
\newblock (2005), hep-ph/0510332.

\bibitem{Dixon:1989fj}
L.~J.~Dixon, V.~Kaplunovsky and J.~Louis,
\newblock Nucl.\ Phys.\ B {\bf 329} (1990) 27.

\bibitem{Bianchi:1991eu}
M.~Bianchi, G.~Pradisi, and A.~Sagnotti,
\newblock Nucl. Phys. {\bf B376}, 365 (1992).

\bibitem{Cremmer:1982en}
E.~Cremmer, S.~Ferrara, L.~Girardello, and A.~Van~Proeyen,
\newblock Nucl. Phys. {\bf B212}, 413 (1983).

\bibitem{Lust:2004fi}
D.~Lust, S.~Reffert, and S.~Stieberger,
\newblock Nucl. Phys. {\bf B706}, 3 (2005), hep-th/0406092.

\bibitem{Lust:2004dn}
D.~Lust, S.~Reffert, and S.~Stieberger,
\newblock (2004), hep-th/0410074.

\bibitem{Font:2004cx}
A.~Font and L.~E. Ibanez,
\newblock JHEP {\bf 03}, 040 (2005), hep-th/0412150.

\bibitem{Jockers:2005zy}
H.~Jockers and J.~Louis,
\newblock Nucl. Phys. {\bf B718}, 203 (2005), hep-th/0502059.

\bibitem{Witten:1985bz}
E.~Witten,
\newblock Nucl. Phys. {\bf B268}, 79 (1986).

\bibitem{Dine:1986zy}
M.~Dine, N.~Seiberg, X.~G. Wen, and E.~Witten,
\newblock Nucl. Phys. {\bf B278}, 769 (1986).

\bibitem{Dine:1987bq}
M.~Dine, N.~Seiberg, X.~G. Wen, and E.~Witten,
\newblock Nucl. Phys. {\bf B289}, 319 (1987).

\bibitem{Berg:2005ja}
M.~Berg, M.~Haack, and B.~Kors,
\newblock (2005), hep-th/0508043.

\bibitem{Abel:2004ue}
S.~A. Abel and B.~W. Schofield,
\newblock JHEP {\bf 06}, 072 (2005), hep-th/0412206.

\bibitem{Abel:2005qn}
S.~A.~Abel and M.~D.~Goodsell,
\newblock  (2005), hep-th/0512072.

\bibitem{Abouelsaood:1986gd}
A.~Abouelsaood, J.~Callan, Curtis~G., C.~R. Nappi, and S.~A. Yost,
\newblock Nucl. Phys. {\bf B280}, 599 (1987).

\bibitem{Dixon:1986qv}
L.~J. Dixon, D.~Friedan, E.~J. Martinec, and S.~H. Shenker,
\newblock Nucl. Phys. {\bf B282}, 13 (1987).

\bibitem{Haggi-Mani:2000uc}
P.~Haggi-Mani, U.~Lindstrom, and M.~Zabzine,
\newblock Phys. Lett. {\bf B483}, 443 (2000), hep-th/0004061.

\bibitem{Marino:1999af}
M.~Marino, R.~Minasian, G.~W. Moore, and A.~Strominger,
\newblock JHEP {\bf 01}, 005 (2000), hep-th/9911206.

\bibitem{Polchinski:1998rq}
J.~Polchinski,
\newblock Cambridge, UK: Univ. Pr. (1998) 402 p.

\bibitem{Billo:2005fg}
M.~Billo, M.~Frau, S.~Sciuto, G.~Vallone, and A.~Lerda,
\newblock (2005), hep-th/0511036.

\bibitem{Cvetic:2001tj}
M.~Cvetic, G.~Shiu, and A.~M. Uranga,
\newblock Phys. Rev. Lett. {\bf 87}, 201801 (2001), hep-th/0107143.

\bibitem{Cvetic:2001nr}
M.~Cvetic, G.~Shiu, and A.~M. Uranga,
\newblock Nucl. Phys. {\bf B615}, 3 (2001), hep-th/0107166.

\bibitem{Cremades:2002te}
D.~Cremades, L.~E. Ibanez, and F.~Marchesano,
\newblock JHEP {\bf 07}, 009 (2002), hep-th/0201205.

\bibitem{DiVecchia:1996uq}
P.~Di~Vecchia, L.~Magnea, A.~Lerda, R.~Russo, and R.~Marotta,
\newblock Nucl. Phys. {\bf B469}, 235 (1996), hep-th/9601143.

\bibitem{Ooguri:1996ck}
H.~Ooguri, Y.~Oz, and Z.~Yin,
\newblock Nucl. Phys. {\bf B477}, 407 (1996), hep-th/9606112.

\bibitem{Dine:1987gj}
M.~Dine, I.~Ichinose, and N.~Seiberg,
\newblock Nucl. Phys. {\bf B293}, 253 (1987).

\bibitem{Billo:2005jw}
M.~Billo, M.~Frau, F.~Lonegro, and A.~Lerda,
\newblock JHEP {\bf 05}, 047 (2005), hep-th/0502084.

\bibitem{Burwick:1990tu}
T.~T. Burwick, R.~K. Kaiser, and H.~F. Muller,
\newblock Nucl. Phys. {\bf B355}, 689 (1991).

\bibitem{Erler:1992gt}
J.~Erler, D.~Jungnickel, M.~Spalinski, and S.~Stieberger,
\newblock Nucl. Phys. {\bf B397}, 379 (1993), hep-th/9207049.

\bibitem{Stieberger:1992bj}
S.~Stieberger, D.~Jungnickel, J.~Lauer, and M.~Spalinski,
\newblock Mod. Phys. Lett. {\bf A7}, 3059 (1992), hep-th/9204037.

\bibitem{Stieberger:1992vb}
S.~Stieberger,
\newblock Phys. Lett. {\bf B300}, 347 (1993), hep-th/9211027.

\bibitem{Seiberg:1999vs}
N.~Seiberg and E.~Witten,
\newblock JHEP {\bf 9909} (1999) 032, hep-th/9908142.

\end{thebibliography}
\end{document}